\documentclass[final,twocolumn,10pt,journal]{IEEEtran}
\usepackage[latin1]{inputenc}

\ifCLASSINFOpdf
  \else
  \fi
\usepackage[cmintegrals]{newtxmath}
\ifCLASSOPTIONcompsoc
\usepackage[caption=false, font=normalsize, labelfont=sf, textfont=sf]{subfig}
\else
\usepackage[caption=false, font=footnotesize]{subfig}
\fi
\usepackage{graphicx}
\usepackage{float}
\usepackage{color} 
\usepackage{epsf}
\usepackage{amsmath, amsfonts,mathtools}
\usepackage{latexsym}
\usepackage{multirow}
\usepackage{multicol}
\usepackage{cite}
\usepackage{algorithm}
\usepackage{algpseudocode}
\usepackage{lipsum}
\usepackage{bm}

\newtheorem{lemma}{\textbf{Lemma}}

\newtheorem{theorem}{\textbf{Theorem}}
\newtheorem{proposition}{\textbf{Proposition}}

\newtheorem{assumption}{\textbf{Assumption}}

\newcommand{\overbar}[1]{\mkern 1.5mu\overline{\mkern-1.5mu#1\mkern-1.5mu}\mkern 1.5mu}
\newcommand{\eqdef}{\vcentcolon=}

\usepackage[cmintegrals]{newtxmath}
\IEEEoverridecommandlockouts

\begin{document}

\title{Wireless Distributed Edge Learning: How Many Edge Devices Do We Need?}

\author{Jaeyoung~Song
	and~Marios~Kountouris,~\IEEEmembership{Senior~Member,~IEEE}
\thanks{Jaeyoung Song and Marios Kountouris are with the Department of Communication Systems, EURECOM, Sophia Antipolis, 06904, France, email: \{jaeyoung.song, marios.kountouris\}@eurecom.fr. Part of this work has been presented in \cite{Song_SPAWC2020}.}
}

\maketitle

\begin{abstract}
We consider distributed machine learning at the wireless edge, where a parameter server builds a global model with the help of multiple wireless edge devices that perform computations on local dataset partitions. Edge devices transmit the result of their computations (updates of current global model) to the server using a fixed rate and orthogonal multiple access over an error prone wireless channel. In case of a transmission error, the undelivered packet is retransmitted until successfully decoded at the receiver.
Leveraging on the fundamental tradeoff between computation and communication in distributed systems, our aim is to derive how many edge devices are needed to minimize the average completion time while guaranteeing convergence. We provide upper and lower bounds for the average completion and we find a necessary condition for adding edge devices in two asymptotic regimes, namely the large dataset and the high accuracy regime.
Conducted experiments on real datasets and numerical results confirm our analysis and substantiate our claim that the number of edge devices should be carefully selected for timely distributed edge learning.
\end{abstract}

\begin{IEEEkeywords}
Distributed machine learning, mobile edge computing, timely computing, computation-communication tradeoff, distributed optimization.
\end{IEEEkeywords}

\IEEEpeerreviewmaketitle

\section{Introduction}\label{sec:introduction}
The ever growing generation and collection of abundant multimodal data, fueled by the interconnection of myriad devices (robots, vehicles, drones, etc.) with advanced sensing, computing, and learning capabilities, will empower emerging applications and automated decision-making. Leveraging on machine learning (ML), data-driven modeling approaches have become ubiquitous in various fields and areas. Scaling up training and inference algorithms, with respect to the number of training examples and/or the number of model parameters, appears to be the prevailing mantra in machine learning as this can drastically improve accuracy. Nevertheless, it is hard for a single entity to efficiently handle high-volume, open-ended datasets. This is exacerbated when data is acquired at the wireless edge and has to be transferred at the central entity. As an alternative, distributed machine learning, which uses multiple devices to collaboratively build a model capitalizing on data or model parallelism, has attracted significant attention \cite{Li_OSDI2014}.
At the same time, many applications, such as self-driving cars and robotic exploration, require timely decision making and fast reactivity. In this context, the process of training an ML model is subject to stringent latency constraints. The emerging edge-aided networking paradigm, which brings communication, computing, and intelligence closer to the network edge and the proximity of end user rather than in the cloud, is a key enabler for near real-time data analytics and low-latency ML \cite{Shi_IOT2016}. In short, distributed edge machine learning is a promising direction for large-scale ML-empowered applications with low latency constraints.

In distributed edge learning, a remote parameter server (PS), which has access to the entire dataset, partitions the training examples into disjoint subsets, which are in turn distributed to multiple edge devices (helpers). The goal is to collaboratively build a global exploiting distributed parallel computing. Since each edge device can access only part of the entire dataset, the outcome of its computation (intermediate results) should be aggregated at the PS to yield a global model. Moreover, since each edge device needs to know the effect of the computation results of other edge devices, information is necessarily sent back to edge devices. In this context, distributed learning entails multiple rounds of communication between the PS and edge devices as a means to produce a suitable model, of  comparable performance to a model built through centralized learning. Several studies have shown that the communication overhead is far from being negligible \cite{Lan_arxiv2019}. In general, the time for communication can be many orders of magnitude longer than the time for local computations \cite{Huang_SIGCOM2013}. Moreover, multiple iterations of information exchange are required for achieving satisfactory accuracy, making the communication between PS and edge devices a dominant factor of the overall performance of distributed learning system \cite{Li_SPM2020}. This motivates the design of communication-efficient distributed learning systems with minimal communication overhead \cite{Li_NIPS2014}.

\subsection{Related Work}
A very popular communication-efficient distributed learning framework is federated learning (FL), in which locally trained models are exchanged instead of datasets \cite{McMahan_MLR2017}. The local models are exchanged or aggregated at the PS to produce a global model. When the PS aggregates local models of computing devices, \cite{Konecny_NIPS2016} proposed receiving local models of only a fraction of devices and averages out the received models to generate global model of next iteration. By randomly choosing a subset of devices, communication overhead can be reduced. Also, when edge devices and the PS exchange gradient in which stochastic gradient descent is used for distributed learning, gradient compression is proposed to reduce communication overhead \cite{Alistarh_NIPS2017, Sun_arxiv2020}. Using compression technique, a distributed FL algorithm was proposed to reduce total communication round in \cite{Mills_IOT2019}. The authors of \cite{Khaled_Arxiv2019} recently studied the iterative compression of model. Using a lossy randomized compression technique, model exchange can be accelerated. In \cite{Amiri_TSP2020}, analog transmission scheme which utilizes the superposition nature of wireless channel was proposed for over-the-air computation. Also, \cite{Ahmed_WSDM2012} considered asynchronous distributed learning algorithm which enables waiting time reduction. In \cite{Caldas_arxiv2018}, a dropout technique for reducing local computation is studied. By using dropout, each device can process on a smaller sub-model. Thus, communication cost can be reduced. For wireless network where nodes can become either the PS or working devices, optimization of forming clusters, which consist of a PS and working devices under constraints of the tradeoff between coverage and computation is shown to be NP-complete \cite{Koyuncu_Arxiv2019}. Also, \cite{Nishio_ICC2019} has  shown that proper user selection based on communication and computation resource can reduce communication overhead. The number of communication rounds required to achieve a target performance has been studied in \cite{Dinh_Arxiv2019, Wang_JSAC2019, Smith_MLR2018}. The work of Dinh \textit{et al.} \cite{Dinh_Arxiv2019} investigated wireless resource allocation problem and solved suboptimally by decomposing the original problem into three sub problems. In \cite{Wang_JSAC2019}, the tradeoff between local update and global parameter aggregation was studied under generic resource constraints. By optimizing the frequency of global aggregation, computation and communication resources are utilized efficiently. Based on duality, distributed learning problem can be solved within arbitrary gap from the optimal solution found via centralized learning \cite{Smith_MLR2018}. The convergence rate is characterized in a closed-form. Utilizing the distributed learning algorithm proposed in \cite{Smith_MLR2018}, the authors of \cite{Yang_TCOM2020} proposed different scheduling policy in wireless networks. Aiming to address communication bottleneck, wireless resource allocation is studied in \cite{Chang_arxiv2020}. By allocating communication resource to devices based on informativeness of the local model and communication channel, performance of federated learning can be improved under fixed communication round. Furthermore, \cite{Chen_arxiv2019} considered transmit power control for transmitting local models to the PS under given resource allocation.

Most prior work on improving communication efficiency of distributed edge learning system have assumed a fixed edge network. Given a network topology, various resource allocation, aggregation, and other computing or communication strategies have been proposed in order to design communication-efficient distributed edge learning system. However, the network topology, i.e., the geographic distribution of edge devices, is not usually optimized. In fact, wireless topology critically affects the performance of distributed edge learning system. Considering that the wireless link quality is highly dependent on phenomena, such as attenuation, shadowing, and fading, which depends on spatial randomness, designing optimal topology for distributed learning becomes important problem to enhance the performance of distributed learning. One parameter that is related to network topology and the network size, is the number of edge devices. Furthermore, the number of edge devices influences both computation and communication delays. On one hand, it is obvious that each edge device is a computing unit that calculates update of model based on local data. In other words, altering the number of edge devices will directly affect the computational efficiency and overall performance of distributed edge learning system. On the other hand, edge devices have communication capabilities for exchanging updates and models with the PS. Since wireless resources are shared by all communication nodes, the total number of edge devices critically determines the amount of wireless resources allocated to each edge device (in orthogonal access) or the amount of interference (in non-orthogonal access). Therefore, it is of cardinal importance to investigate the effect of the number of edge devices on the performance of communication-efficient distributed edge learning.

\subsection{Contributions}
The objective of this work is to derive the optimal number of edge devices in a distributed edge learning system operating under latency constraints. If an edge device is added to the system, computing load will be shared; thus, the computation time per device will decrease. In contrast, an additional edge device will have to be allocated wireless resources to transmit its intermediate results. However, since wireless resources are limited and do not usually scale with the number of participating devices, as the number of edge devices increases, available resources per edge device are reduced. Also, in non-orthogonal multiple access, higher interference is generated when more edge devices transmit to the PS simultaneously. As a consequence, increasing the number of edge devices can lead to longer communication time. By investigating the tradeoff between computation and communication as a function of edge devices, we aim at minimizing the completion of the distributed learning process while guaranteeing a certain accuracy and convergence. When transmission errors occur, the undelivered packets are retransmitted until successfully decoded at the receiver, either edge devices or PS. 

Taking wireless channel into account, we examine how many edge devices are necessary to achieve minimum completion time. Furthermore, the upper bound and lower bound of completion time are characterized in a close-form. Using closed-form expression, we can find necessary condition on the number of edge devices in asymptotic regime. Also, we present experiments and numerical results to obtain further insights on optimal number of edge devices.
In short, our contributions are summarized as follows.
\begin{itemize}
	\item We formulate the problem of minimizing the average completion time which is defined as time duration from the beginning of the the distributed learning process until obtaining the final output considering both computation and over-the-air communication aspects.
	\item For analytical tractability, we derive in closed-form an upper and a lower bound for the average completion time.
	\item In the high accuracy regime, by comparing the upper and lower bounds, we provide the necessary condition for adding an edge device without deteriorating the completion time performance of the system.
	\item As the upper bound tightly approximates the average completion time for small number of edge devices, we consider the minimization of the upper bound on the completion time in the very large data regime. A necessary condition for optimality is also provided as a function of the number of devices and the minimum average received SNR among edge devices.
\end{itemize}

\subsection{Organization}
The remainder of the paper is organized as follows. In Section \ref{sec:system_model} we introduce the system model and in Section \ref{sec:problem_formulation}, the completion time minimization problem is formulated. 
The analysis for the optimal number of edge devices is given in Section \ref{sec:analysis}. Experiments and numerical results are provided in Section \ref{sec:simulations}, and Section \ref{sec:conclusions} concludes our work.

\section{System Model}\label{sec:system_model}

We consider a distributed edge learning system with a single PS and a set of $K$ edge devices, represented by $\mathcal{K} = \left\lbrace 1,\cdots K \right\rbrace$. We describe below the learning framework and the system model. 
\begin{figure} 
	\centering
	\subfloat[Data distribution\label{2a}]{%
		\includegraphics[scale=0.15]{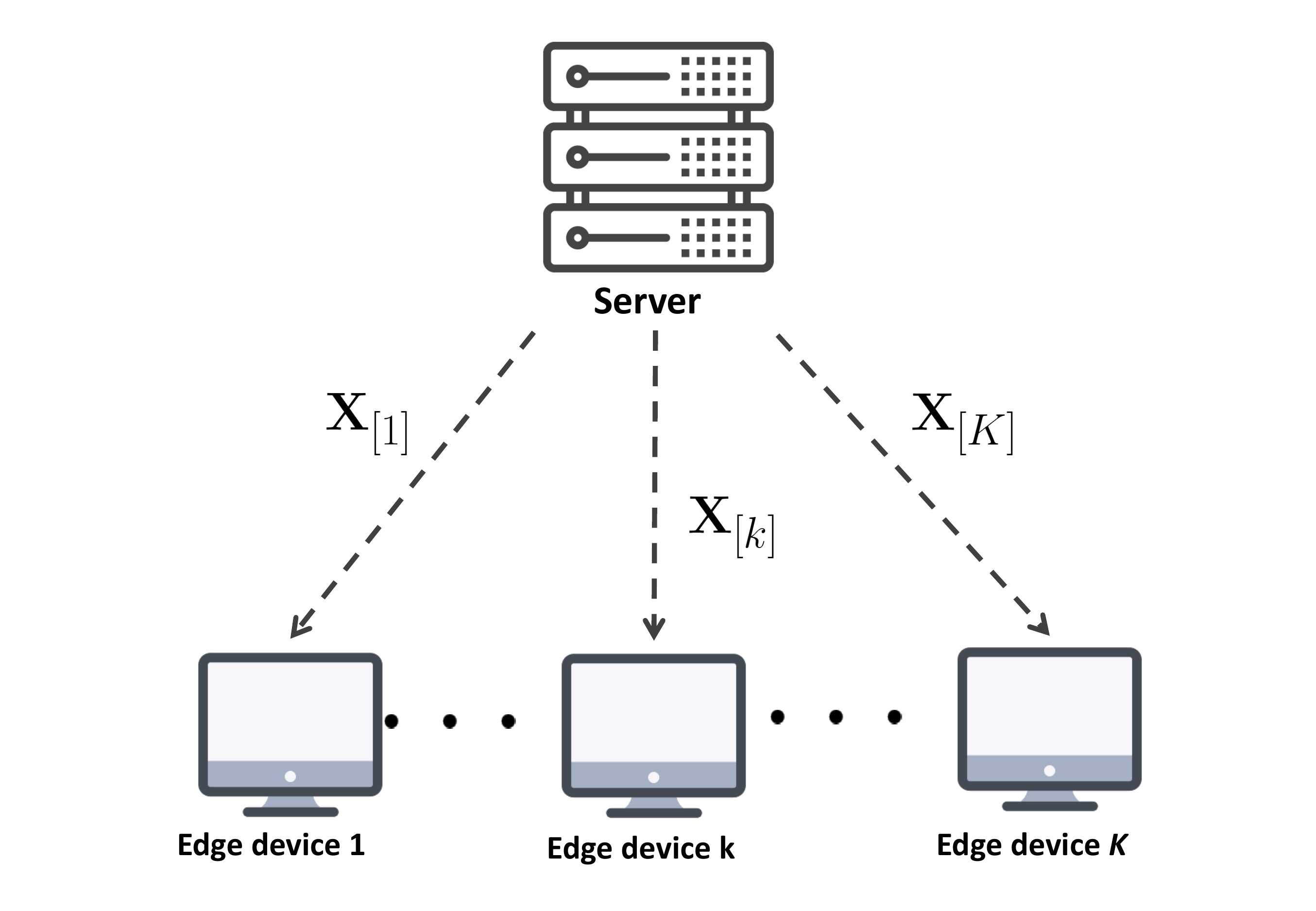}}
	\subfloat[Local computation\label{2b}]{%
		\includegraphics[scale=0.15]{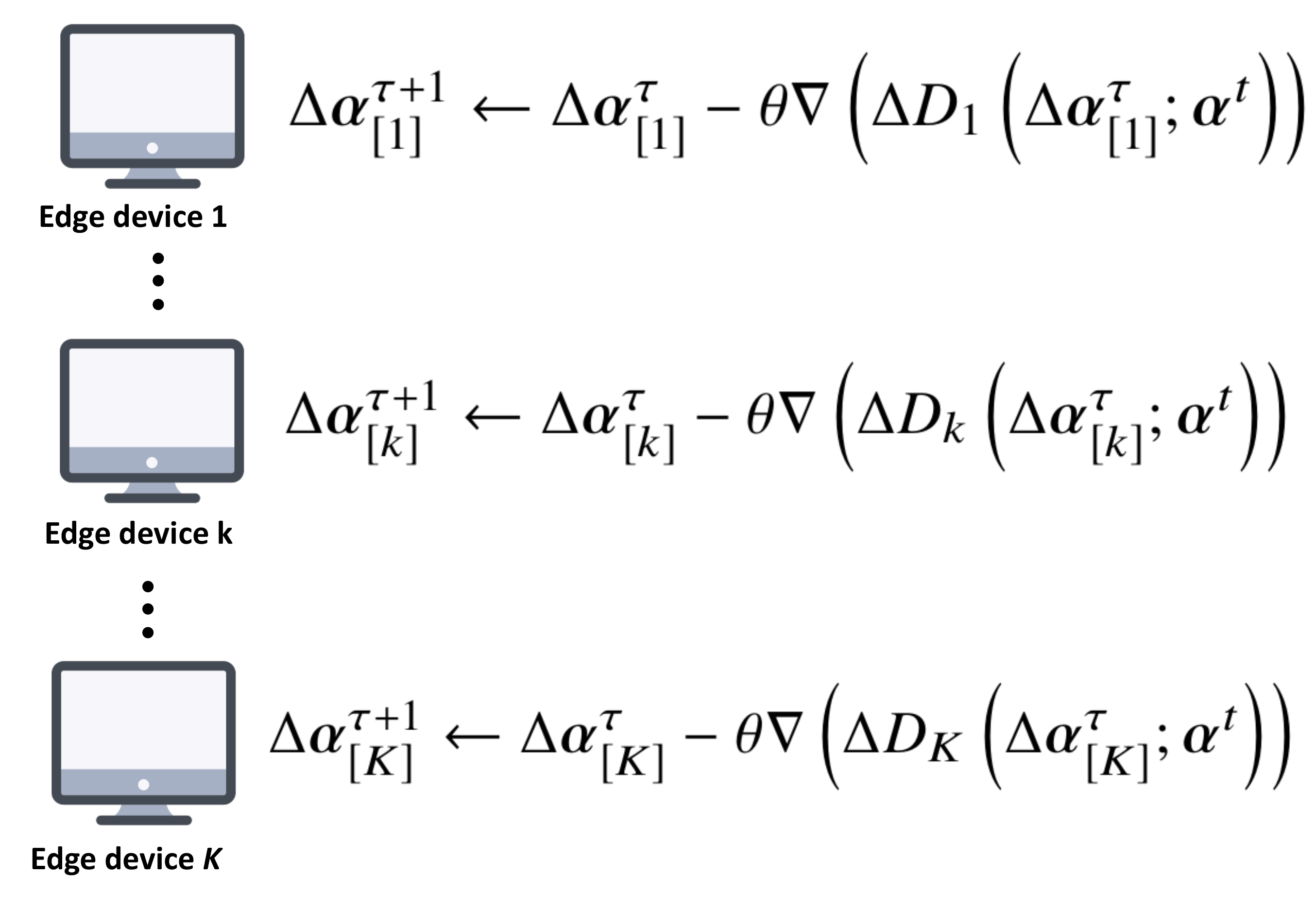}}\\
	\subfloat[Local update delivery\label{2c}]{%
		\includegraphics[scale=0.15]{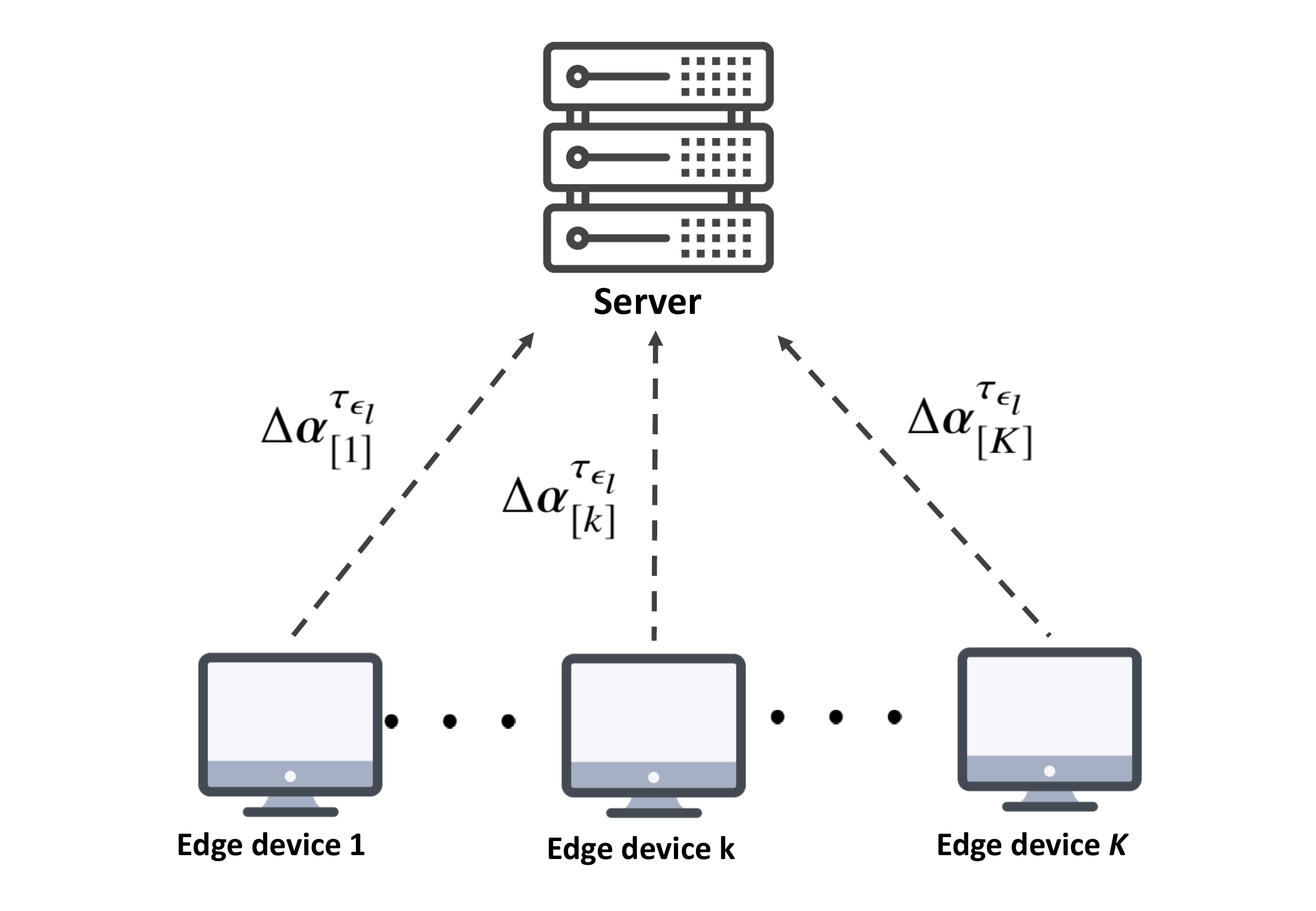}}
	\subfloat[Global model delivery\label{2d}]{%
		\includegraphics[scale=0.15]{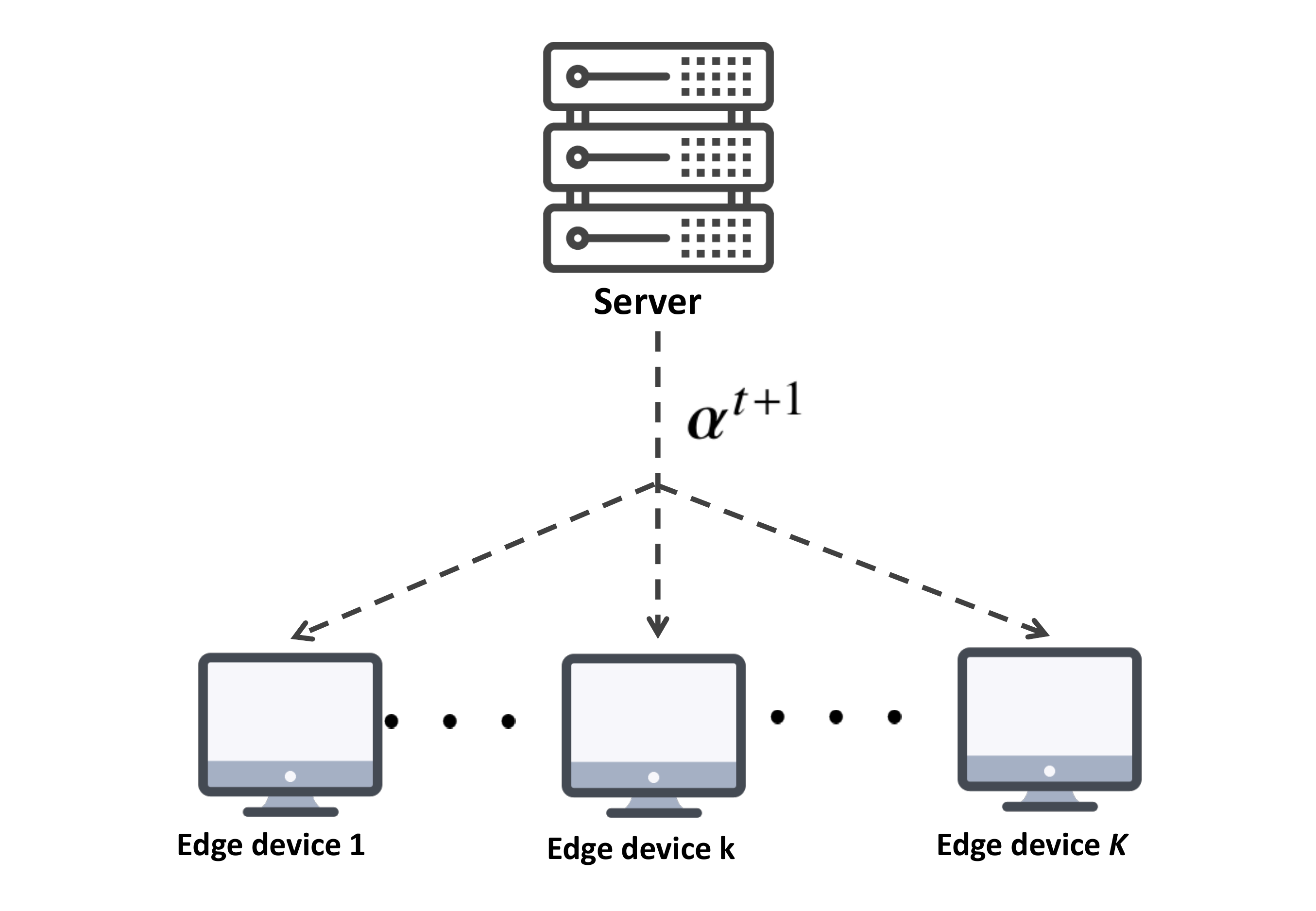}}
	\caption{Procedures in distributed learning}
	\label{fig:system_model_2} 
\end{figure}

\subsection{Distributed Learning}
In our setting, the PS holds the entire dataset and aims to find and train a global model that fits to this dataset. Leveraging on parallel edge computing, the PS distributes different parts of the dataset to multiple edge devices. After data distribution, each edge device produces local updates based on the partial dataset it has access to and the current global model received from the PS. Edge devices send their local updates of current global model, once computed, back to the PS. The PS gathers local updates from all edge devices and calculates an updated, normally improved, global model. The updated global model is communicated to edge devices and is in turn used for calculating next local updates. The procedures (phases) that entail local updates computation, transmission of local updates, and updated global model broadcasting, are repeated until the global model achieves a target performance. As shown in the Figure \ref{fig:system_model_2}, in each global iteration, local computation, local update delivery, and global model delivery is repeated. Note that in every global iteration, local data is not exchanged but models are transferred. In addition to this, our learning framework converges under heterogeneous data distribution; the results and analysis of this paper on the average completion time can be extended to federated learning.

Suppose we have a set of $N$ data examples represented by a matrix $\mathbf{X} = \left[ \mathbf{x}_1, \cdots, \mathbf{x}_N \right] \in \mathbb{R}^{M \times N}$, where $\mathbf{x}_n \in \mathbb{R}^{M}$ is the $n$-th data example characterized by $M$ features. As the error caused by approximating expected risk with empirical risk is bounded by $\mathcal{O}\left( \sqrt{\frac{\log N}{N}} \right)$ \cite{Bottou_COMS2010}, we define our objective function $F(\mathbf{w})$ for a model (parameter) $\mathbf{w} \in \mathbb{R}^M$ as the regularized empirical risk, i.e.,
\begin{align}
F(\mathbf{w}) =  \frac{1}{N}\sum_{n=1}^N \ell_n \left( \mathbf{x}_n^\top \mathbf{w} \right) + \lambda r(\mathbf{w}), \label{eq:primal_obj}
\end{align} where $\ell_n\left(\cdot \right)$ is loss function associated with an observation $\mathbf{x}_n$, $r(\cdot)$ is a regularizer, such as $\ell_1$ or $\ell_2$ norm, and $\lambda$ is the weight of the regularization term. The goal of the server is to find the global model that is a solution to the regularized empirical risk minimization problem, i.e., the parameter that minimizes $F(\mathbf{w})$
\begin{align}
\min_{\mathbf{w} \in \mathbb{R}^d} F(\mathbf{w}). \label{eq:orig_obj}
\end{align}

Distributed edge computing is employed to overcome the difficulty in handling very large datasets, i.e., the server builds the global model by coordinating multiple edge devices. First, the server splits the dataset into $K$ subsets, one for each edge device. Let us denote a set of indexes of data examples allocated to edge device $k$ as $\mathcal{P}_k \subseteq \left\lbrace 1, \cdots, N \right\rbrace$. Thus, $\left\lbrace \mathcal{P}_k \right\rbrace_{k=1}^K$ forms a partition of the entire dataset. 
We assume that duplicate allocation of a data example is not allowed and all examples should be given to at least one edge device, i.e., $\mathcal{P}_k \cap \mathcal{P}_{k'} = \emptyset$ and $\cup_{k=1}^K \mathcal{P}_k = \left\lbrace 1, \cdots, N \right\rbrace$). The number of data examples assigned to edge device $k$ is represented by $n_k$, with $\sum_{k=1}^K n_k = N$.

\begin{algorithm}[t!]
	\caption{CoCoA-Distributed Learning Algorithm}
	\label{alg:cocoa}
	\begin{algorithmic}
		\Require Initial point $\bm \alpha^0$, aggregation parameter $\gamma$, partition of the entire data set $\left\lbrace \mathcal{P}_k \right\rbrace_{k=1}^K$
		\For{$t=0,1,2,\cdots,$}
		\For{$k \in \left\lbrace 1,2,\cdots,K \right\rbrace$ each edge device $k$ does in parallel}
		\For{$\tau = 1,2, \cdots, \tau_{\epsilon_l} - 1 $}
		\State $\Delta \bm \alpha_{[k]}^{\tau+1} \gets \Delta \bm \alpha_{[k]}^{\tau} - \theta \nabla \left( \Delta D_k \left( \Delta \bm \alpha_{[k]}^\tau ; \bm\alpha^t \right) \right)$
		\EndFor
		\State Transmit $\Delta \bm \alpha_{[k]}^{\tau_{\epsilon_l}}$ to the server
		\EndFor
		\State Produce global parameter: $\bm \alpha^{t+1} = \bm \alpha^t + \gamma \sum_{k=1}^K \Delta \bm \alpha_{[k]}^{\tau_{\epsilon_l}}$
		\State Multicast $\mathbf{X} \bm \alpha^{t+1}$ to all edge devices.
		\EndFor		 
	\end{algorithmic}
\end{algorithm}

Given a data partition, the PS utilizes distributed learning framework to efficiently solve \eqref{eq:orig_obj} with a large set of data examples.  Among several existing distributed learning frameworks \cite{Wang_JSAC2019, Smith_MLR2018, Dinh_Arxiv2019}, we choose to employ the widely used communication-efficient distributed dual coordinate ascent (CoCoA) framework, which guarantees convergence \cite{Smith_MLR2018}. However, our analysis can be readily extended to other learning scheme such as \cite{Wang_JSAC2019, Dinh_Arxiv2019} by replacing the number of iteration required for given learning framework to achieve optimality gap. Since duality is used in CoCoA distributed learning framework, the goal is to find global parameter $\bm\alpha \in \mathbb{R}^M$ which is related with model $\mathbf{w}$ via dual relationship. The global model $\mathbf{w}$ is related with global parameter $\bm \alpha$ via $\mathbf{w} = \nabla r^* \left( \frac{1}{\lambda N} \mathbf{X} \bm\alpha\right) $, where $r^*(\cdot)$ is convex conjugate of $r(\cdot)$. 

The CoCoA framework starts from data distribution. First, $\mathbf{X}_{[k]} \in \mathbb{R}^{M \times N}$, defined as a matrix whose column vectors are data examples allocated to edge device $k$, is distributed among devices, i.e., $\left(\mathbf{X}_{[k]} \right)_{ij} = \left(\mathbf{x}_j\right)_i$ if $i \in \mathcal{P}_k$, and $\left(\mathbf{X}_{[k]} \right)_{ij} = 0$, otherwise. Then, each edge device locally computes updates $\Delta \bm\alpha_{[k]}$ for given global parameter $\bm\alpha$ by solving local subproblem (\ref{prob:local}) defined below. The $k$-th edge device transmits $\Delta \bm\alpha_{[k]}$ to the PS. After receiving $\Delta \bm\alpha_{[k]}$ from each edge device, the PS updates the global parameter $\bm\alpha$ and delivers updated global parameter $\bm\alpha$ to every edge device. We define a global iteration as a round of local computation and transmission from edge devices and from the PS. The initial data distribution and algorithmic sequence of $t$-th global iteration are shown in Figure \ref{fig:system_model_2}. The local subproblem that edge device $k$ solves can be written as
\begin{align}
\min_{\Delta \bm\alpha_{[k]} \in \mathbb{R}^N} \Delta D_k \left( \Delta \bm\alpha_{[k]}; \bm\alpha^t\right) \label{prob:local}
\end{align} where 
	\begin{align}
	\nonumber &\Delta D_k(\Delta \bm \alpha_{[k]}^t; \mathbf{X} \bm\alpha) = \frac{\lambda}{K} r^*\left(\frac{1}{\lambda N} \mathbf{X} \bm\alpha^t \right) +  \frac{1}{N} \mathbf{w}^\top \left( \mathbf{X} \Delta \bm\alpha_{[k]} \right) \\
	&+ \frac{\gamma \sigma'}{2 \lambda N^2} \left|\left| \mathbf{X}\Delta \bm\alpha_{[k]} \right|\right|^2 + \frac{1}{N} \sum_{n \in \mathcal{P}_k} \ell_n^*\left( - \alpha_n^t - \left(\Delta \bm\alpha_{[k]}\right)_n \right), 
\end{align} where $\sigma'$ is defined as
\begin{align}
\sigma' &= \frac{1}{K} \max_{\bm\alpha} \frac{|| \mathbf{X} \bm \alpha ||^2}{\sum_{k=1}^K || \mathbf{X}_{[k]} \bm\alpha_{[k]}||^2}.
\end{align} and $r^*(\cdot)$ and $\ell_n^*(\cdot)$ are convex conjugate functions of $r(\cdot)$ and $\ell_n(\cdot)$, respectively. Also, $\gamma$ is an aggregation weight, which controls how the updates from each edge device are combined, $\Delta \bm\alpha_{[k]}$ is the local update of edge device $k$, and $\bm \alpha^t$ is the global parameter at the $t$-th global iteration. The details of CoCoA distributed learning framework is represented in Algorithm \ref{alg:cocoa}.

Since CoCoA is based on duality, its convergence is determined by duality gap defined as $G(\bm\alpha^t) = F(\mathbf{w}(\bm\alpha^t)) - D(\bm\alpha^t)$, where $D(\bm\alpha^t)$ is a dual function of $F(\mathbf{w})$. Note that it can be proven that optimality gap is smaller than duality gap \cite{Boyd_book2004}. Thus, a solution satisfying duality gap is always within optimality gap from the optimal solution. If we repeat the procedures of local computation and exchange of the local updates and global parameters, after a finite number of iterations, the following duality gap can be obtained:

Suppose the loss function and the regularizer satisfy the following assumption, respectively.
\begin{assumption}
	The loss function $\ell(\cdot)$ is a $\frac{1}{\mu}$-smooth function satisfying that for $\forall \mathbf{u}, \mathbf{v} \in \mathbb{R}^M$,
	\begin{align}
	\ell(\mathbf{u}) \leq \ell(\mathbf{v}) + \nabla \ell(\mathbf{v})^\top \left( \mathbf{u} - \mathbf{v} \right) + \frac{1}{2\mu} \left|\left| \mathbf{u} - \mathbf{v} \right|\right|^2 \label{ineq:smooth}
	\end{align}
\end{assumption}
\begin{assumption}
	The regularizer $r(\cdot)$ is a $\zeta$-strongly convex function that satisfies for $\forall \mathbf{u}, \mathbf{v} \in \mathbb{R}^M$
	\begin{align}
	r(\mathbf{u}) \geq r(\mathbf{v}) + \nabla r(\mathbf{v})^\top \left( \mathbf{u} - \mathbf{v} \right) + \frac{\zeta}{2} \left|\left| \mathbf{u} - \mathbf{v} \right|\right|^2 \label{ineq:convexity}
	\end{align} 	
\end{assumption}
\begin{theorem}[Theorem 11 in \cite{Smith_MLR2018}]\label{thm:convergence}
	Suppose we have $\epsilon_l$-accuracy solution of local subproblem.
	For $\frac{1}{\mu}$-smooth loss function and $\zeta$-strongly convex regularizer, given $\mathbf{X}$ and $\mathbf{X}_{[n]}$, $\forall k \in \mathcal{K}$, if $t \geq M_K $,
	\begin{align}
		G(\alpha^t) \leq \epsilon_G,
	\end{align} where
	\begin{align}
		M_K = \left\lceil \frac{K}{ \left( 1 - \epsilon_l \right)} \frac{ \mu \zeta \lambda N  + \sigma' \sigma_{\max} }{\mu \zeta \lambda N} \ln\left( \frac{   \lambda \zeta \mu N  + \sigma'\sigma_{\max} }{ \left( 1 - \epsilon_l \right)\lambda \zeta \mu N } \cdot \frac{K}{\epsilon_G} \right) \right\rceil , \label{eq:time_dualgap}
	\end{align} where $\sigma_{\max}$ is defined as
	\begin{align}		
		\sigma_{\max} & = \max_{k} \max_{\bm\alpha_{[k]}} \frac{||\mathbf{X}_{[k]} \bm\alpha_{[k]} ||^2}{||\bm\alpha_{[k]}||^2}.
	\end{align}	
\end{theorem}
The above result states that we can find a sequence of solutions resulting from Algorithm \ref{alg:cocoa}, which reduces the dual function exponentially. As a result, the number of iterations required to satisfy a given duality gap is a logarithmic function of the duality gap. In addition to that, in general, the number of global iteration required to achieve given duality gap increases as the number of edge devices increases. This is because the solution of the local subproblem can be partially aligned with the gradient direction based on entire data set when the local subproblem is defined with small amount of data. Thus, more global iterations are required when data allocated to each edge devices is small, equivalently, when the number of edge devices is large.

\subsection{Communication Model}
In order to find optimal solution in a distributed manner, communication between edge devices and the PS is necessary. We consider here that edge devices and the PS transmit the required information over the air, and we assume perfect channel state information (CSI) at the receivers and absence of CSI at the transmitters (CSIT). As a result, transmitters transmit with fixed rate, which is determined so as to satisfy the application requirements. Moreover, since link capacity is a function of wireless channel, which is randomly varying, communication may be unsuccessful, i.e., a transmission error can occur. This error event is dominated by the outage event defined as the event that the capacity is lower than the predefined fixed transmission rate \cite{Tse_FW2005}. Consequently, retransmission is necessary when outage occurs. In this work, we assume that the transmitters (PS and edge devices) retransmit the undelivered packet until it is successfully decoded at the receiver. In the CSIT case, the transmission rate can be adapted to the instantaneous channel quality state, thus avoiding transmission errors. Consequently, average completion time when CSIT is available may be significantly different from that of in absence of CSIT. Investigating average completion time under CSIT is beyond the scope of this work.

When the PS distributes the data in the beginning of the distributed learning process, edge devices are supposed to receive different subsets of data. Hence, orthogonal bandwidth allocation is considered in order to avoid interference. According to a given bandwidth and transmit power strategy, the PS allocates bandwidth and transmit power to edge devices. Let us denote bandwidth and transmit power to edge device $k$ as $B_k$ and $P_{k}^{\text{Ps}}$, respectively. Also, we assume that the PS have $B$ total bandwidth and $P^{\text{PS}}$ transmit power. Then, the outage probability can be written as
\begin{align}
	p_{k|K}^{\text{dist}} = \mathbb{P} \left[ B_k \log \left( 1 + \frac{g_k P_k^{\text{PS}}}{B_k N_0} \right) < R^{\text{dist}} \right], \label{eq:outage}
\end{align} where $g_k$ denote the wireless channel gain, $N_0$ is the noise power spectral density and $R^{\text{dist}}$ is the predefined transmission rate of each data example during data distribution. After some manipulations, we can rewrite \eqref{eq:outage} as,
\begin{align}
	p_{k|K}^{\text{dist}} = \mathbb{P} \left[ \rho_k < \theta_k \right], 
\end{align} where $\rho_k =\frac{g_k P^{\text{PS}}_t}{B N_0}$ and $\theta_k = \frac{P^{\text{PS}}B_k}{P_{k}^{\text{PS}}B} \left( 2^{\frac{R^{\text{dist}}}{B_k}} - 1 \right)$.

To investigate the relationship between the outage probability and the number of edge devices, we need to specify bandwidth and transmit power allocation scheme. As the PS does not have CSI, existing resource allocation scheme which requires CSI cannot be applied in our work. Instead, uniform bandwidth and power allocation is widely used when CSI is not available at transmitters \cite{Palomar_TIT2003}. Thus, in our paper, we assume uniform resource allocation. Thus, henceforth, $B_k = \frac{B}{K}$ and $P_k^{\text{PS}} = \frac{P_k^{\text{PS}}}{K}$. However, we can readily extend our analysis to non-uniform resource allocation by modifying $\theta_k$ for a certain resource allocation scheme.

When each edge transmits its local update to the PS, multiple edge devices simultaneously transmit signals to single receiver. For tractability, we assume orthogonal multiple access (OMA) during this phase. The effect of non orthogonal multiple access (NOMA) is investigated in Section \ref{sec:simulations}. The link capacity of edge device $k$ in local update delivery under uniform bandwidth allocation can be written as 
\begin{align}
	C_{k|K}^{\text{OMA}} = \frac{B}{K} \log \left( 1 + K \frac{g_k P^{\text{device}}}{BN_0} \right) \label{eq:capacity_oma}
\end{align} where $P^{\text{device}}$ is transmit power of edge devices.
Note that each edge device transmits accessing a fraction of the total bandwidth; hence, noise power, which is proportional to bandwidth, becomes smaller as the allocated bandwidth decreases. 
In data transmission, transmit power is also shared and for each edge device, it becomes the same fraction of bandwidth allocated. Thus, noise power reduction cancels out transmit power reduction in received SNR, which makes that the received SNR is not a function of the total number of edge devices. However, in the multiple access channel, transmit power does not have to be shared among edge devices, and is considered fixed. Therefore, the received SNR increases due to noise reduction as the number of edge devices increases given uniform bandwidth allocation.

Using \eqref{eq:capacity_oma}, the outage probability during local update delivery can be expressed as
\begin{align}
	p^{\text{up}}_{k|K} = \mathbb{P} \left[\frac{B}{K} \log \left( 1 + K \eta_k \right) < R^{\text{up}}\right]
\end{align} where $R^{\text{up}}$ is a given transmission rate for local update and $\eta_k = \frac{g_k P^{\text{device}}}{B N_0}$.

Upon receipt of the local updates from all edge devices, the PS produces a global model for next global iteration. The updated global model is sent back over-the-air to edge devices. Since information sent during data distribution and local update delivery is distinct for each edge device, independent messages should be sent to PS. In other words, unicast transmission is required during those phases. In contrast, the global model is common information to be received by all edge devices. Thus, multicast transmission is used instead, over the whole available bandwidth. The multicast capacity is determined by the link capacity of the worst receiver to ensure that all users can decode their packets successfully. Thus, the rate during the global model delivery phase is given by
\begin{align}
	C^{\text{mul}}_K = B \log \left( 1 + \min_k \frac{g_k P^{\text{PS}}}{B N_0} \right). 
\end{align}
As a result, the outage probability in global model delivery given $K$ edge devices is represented as
\begin{align}
	p^{\text{mul}}_K = \mathbb{P} \left[B \log \left( 1 + \min_k \rho_k \right) < R^{\text{mul}}\right]
\end{align} 
where $R^{\text{mul}}$ is transmission rate for global model transmission and $\min_k \rho_k$ is the minimum received SNR among $K$ edge devices.

\section{Completion Time Minimization in Wireless Edge Learning}\label{sec:problem_formulation}
Evidently, local computing load decreases as the number of edge devices increases, due to parallel distributed processing at the edge devices. Nevertheless, as the number of edge devices increases, communication overhead, mainly exchanging local updates and global model, becomes larger. Hence, there exists a tradeoff between exploiting parallel distributed computing and reducing communication overhead. In this section, we aim at deriving the optimal number of edge devices, which provides the optimal computation-communication tradeoff by minimizing the completion time defined as time taken to achieve certain duality gap of our objective function.

To derive the completion time, we first need to characterize the time taken by each procedure in the distributed learning algorithm. The distributed learning algorithm consists of four different procedures (phases). The first phase is the distribution of partitioned data to each edge device. The PS delivers a part of data set to each edge device over a wireless channel. Let us assume that a single transmission can deliver a single data example. In other words, the transmission rate in data distribution is equal to the size of the single data example. Since $n_k$ data examples are given to edge device $k$, $n_k$ transmissions are required to deliver all data to the $k$-th edge device in the absence of transmission errors. 
However, errors may occur when link capacity becomes lower than transmission rate due to wireless channel randomness. In case of unsuccessful reception, we consider that a retransmission takes place. Let us denote $L^{\text{dist}}_{k|K}$ a random variable that represents the number of transmissions required for delivering a single data example to edge device $k$ given $K$ edge devices. Assuming independent and identically distributed (i.i.d.) channel realizations at each transmission attempt, the distribution of $L^{\text{dist}}_{k|K}$ is identical for any data example. Thus, the time to transmit $n_k$ data examples to edge device $k$ can be represented as
\begin{align}
	T^{\text{dist}}_{k|K} = \omega n_k L^{\text{dist}}_{k|K}
\end{align} where $\omega$ is time duration of a single transmission.

Since data distribution phase is completed when all $K$ edge devices have received their local data successfully, the time required for data distribution is given by the longest time among edge devices, i.e., 
\begin{align}
	T^{\text{dist}}_K = \max_{k \in \mathcal{K}} T_{k|K}^{\text{dist}}
\end{align}

Upon receipt of data examples, each edge device solves the local subproblem to find local update at each global iteration. We assume that edge devices utilize gradient descent (GD) to solve its local subproblem. It is known that the number of iterations required to have $\epsilon_l$ accuracy with GD is given by $O\left(\frac{1}{\epsilon_l} \right)$ \cite{Boyd_book2004}. Furthermore, the processing time of edge device for calculating the gradient at every local iteration is proportional to the number of data examples given to edge devices. Hence, the time for local computation of edge device $k$ is given by
\begin{align}
	T^{\text{local}}_{k|K} = c_k \frac{n_k}{\epsilon_l}
\end{align} where $c_k$ is a constant related to $k$-th edge device's computational capability.

Depending on the number of data examples each edge device processes and its computational capability, the time to find a solution to the local subproblem is different. We define the time for local computing as the maximum of $T^{\text{local}}_k$, i.e., 
\begin{align}
	T^{\text{local}}_K = \max_{k \in \mathcal{K}} T^{\text{local}}_k. \label{eq:time_local}
\end{align}

The local updates obtained by solving local subproblem are then sent to the PS in order to update the global model for the next (global) iteration. Since a local update is a vector whose size is comparable to the size of a single data example, local updates can be carried by a single transmission. Therefore, if we denote $L^{\text{up}}_{k|K}$ the number of transmission for delivering the local update of edge device $k$, the time for local update delivery of edge device $k$ becomes
\begin{align}
	T^{\text{up}}_{k|K} = \omega L^{\text{up}}_{k|K}.
\end{align}
To guarantee convergence of distributed learning, all local updates from $K$ edge devices are required at every global iteration. Therefore, the time to receive all $K$ local updates is obtained by taking the maximum as follows
\begin{align}
	T^{\text{up}}_K = \max_{k \in \mathcal{K}} T^{\text{up}}_{k|K}.
\end{align} 

Upon receipt of local updates from $K$ edge devices, the global model for the next global iteration can be updated according to Algorithm \ref{alg:cocoa}. The PS sends the updated global model to all edge devices to enable them solve their local subproblems for the next global iteration. Since the global model is common information, a multicast transmission, which delivers (broadcasts) a common message to $K$ edge devices with a single transmission, can be employed. Let $L^{\text{mul}}_K$ be the number of transmissions required for multicast transmission of the global model to $K$ edge devices, the time for accomplishing the global model delivery phase is given by
\begin{align}
	T^{\text{mul}} = \omega L^{\text{mul}}_K. \label{eq:time_mul}
\end{align}

In this paper, we assume that each procedure starts in a synchronized manner across all edge devices. Synchronous algorithm guarantees all edge devices have the equivalent model which is important to prove the convergence. Also, it is relatively simple and easy to implement than asynchronous model. However, synchronous algorithm is susceptible to stragglers. While, asynchronous algorithm can mitigate stragglers by overlapping communications and computations \cite{Li_SPM2020}. The comparison of synchronous and asynchronous algorithms are beyond the scope of this paper. For tractability of analysis, synchronized learning is used. Thus, a next phase can start when all edge devices have accomplished the current phase. In addition to that, from Theorem \ref{thm:convergence}, we need $M_K$ global iterations to guarantee that our global model satisfies a predefined duality gap. Consequently, the completion time of the distributed learning with $K$ edge devices is obtained as
\begin{align}
	T^{\text{DL}}_K &= T^{\text{dist}}_K + M_K \left( T^{\text{local}}_K + T^{\text{up}}_K + T^{\text{mul}}_K \right).
\end{align} 
Note that the time for each procedure at each global iteration is i.i.d given that channel realizations at each transmission attempt are i.i.d.
Moreover, data distribution procedure is performed only once at the very beginning of the training process, whereas all other procedures are repeated at every global iteration.

Our goal is to minimize the average completion time of distributed edge learning. This optimization problem can be formulated as follows.
\begin{align}
	&\min_{K} \overbar{T^{\text{DL}}_K} \\
	\text{s.t. } & K \in \mathbb{Z}^+,
\end{align} where $ \overbar{T^{\text{DL}}_K} = \mathbb{E}\left[ T^{\text{DL}}_K \right]$ and $\mathbb{Z}^+$ is a set of positive integer.

\section{Optimal Number of Edge Devices}\label{sec:analysis}

In this section, we analyze the optimal number of edge devices that minimizes the average completion time of distributed edge learning. The wireless channel is subject to Rayleigh fading, i.e., the channel power gain follows an exponential distribution. However, distribution of wireless channel for edge devices can have different mean due to different location and propagation environment. Under these assumptions, the outage probability for the data distribution and the local update delivery phase can be characterized as follows, respectively.
\begin{align}
p^{\text{dist}}_{k|K} &= 1 - \exp \left(-\frac{1}{\overbar{\rho}_k}\left(2^{\frac{KR^{\text{dist}}}{B}} - 1 \right) \right), \label{eq:outage_dist}\\
p^{\text{up}}_{k|K} &= 1 - \exp \left(-\frac{1}{K\overbar{\eta}_k}\left(2^{\frac{KR^{\text{up}}}{B}} - 1 \right) \right),	\label{eq:outage_up}
\end{align} 
where $\overbar{\rho}_k = \mathbb{E}\left[ \rho_k \right]$ and $\overbar{\eta}_k = \mathbb{E}\left[ \eta_k \right]$.
Using \eqref{eq:outage_dist} and \eqref{eq:outage_up}, we can derive the probability mass function (PMF) of the number of transmissions in data distribution and local update delivery. Suppose the PS transmits $l$ times for delivering a single data example to edge device $k$. This event is equivalent to the event that $l-1$ outages occurred from the first transmission attempt until $(l-1)$-th transmission attempt and transmission is successful at the $l$-th transmission attempt. Therefore, we have the following PMF of the number of transmissions to edge device $k$ in data distribution $L^{\text{dist}}_{k|K}$.
\begin{align}
	\mathbb{P} \left[ L^{\text{dist}}_{k|K} = l \right] = \left( p^{\text{dist}}_{k|K} \right)^{l-1} \left(1 - p^{\text{dist}}_{k|K} \right). \label{eq:PMF_L_dist}
\end{align}

The average completion time, using the linearity of expectation, can be expressed as
\begin{align}
	\overbar{T^{\text{DL}}_K} &= \mathbb{E}\left[ T^{\text{dist}}_K \right] + \mathbb{E}\left[ M_K \left( T^{\text{local}}_K + T^{\text{up}}_K + T^{\text{mul}}_K \right)\right], \\
	\nonumber &= \omega \mathbb{E}\left[ \max_{k \in \mathcal{K}}n_k L^{\text{dist}}_{k|K} \right] + M_K \frac{\max_k\left\lbrace c_k n_k \right\rbrace}{\epsilon_l} \\
	&+ M_K \omega \mathbb{E}\left[ \max_{k \in \mathcal{K}} L^{\text{up}}_{k|K} \right] + M_K \omega \mathbb{E}\left[ L^{\text{mul}}_K \right]. \label{eq:completion_time}
\end{align}
Let $\max_{k \in \mathcal{K}} L_{k|K}^{\text{dist}}$ be the largest number of transmissions among $K$ edge devices in data distribution; the probability distribution of $L^{\text{dist}}_{k|K}$ is not identical for different $k$. Although the type of distribution for wireless channel gain is identical as Rayleigh, depending on the average received SNR, $\overbar{\rho}_k$, the outage probability $p^{\text{dist}}_{k|K}$ becomes different; thus, $L^{\text{dist}}_{k|K}$ has different distribution for different $k$. Unfortunately, the distribution of $\max_{k \in \mathcal{K}} L_{k|K}^{\text{dist}}$, which corresponds to order statistics with non-identically distributed random variables, is not tractable. However, we resort to the following lower and upper bounds of the average completion time using order statistics and considering the worst and best cases for the average received SNR.
\begin{proposition}\label{prop:bounds}
	The average completion time of the distributed edge learning system with $K$ edge devices is bounded as
	\begin{align}
		\overbar{T^{\text{DL}}_{\min|K}} \leq \overbar{T^{\text{DL}}_{K}} \leq \overbar{T^{\text{DL}}_{\max|K}} \label{ineq:relation_t_dl}
	\end{align} 
	where
	\begin{align}
		\nonumber \overbar{T^{\text{DL}}_{\max|K}} & = \omega \max_k \left\lbrace n_k \right\rbrace \sum_{q=1}^K {K \choose q} \frac{\left( -1 \right)^{q+1}}{ 1 - \left(p^{\text{dist}}_{\max|K} \right)^q } + M_K  \frac{\max_k \left\lbrace c_k n_k \right\rbrace }{\epsilon_l}\\
		& + \omega M_K \sum_{q=1}^K {K \choose q}  \frac{\left( -1 \right)^{q+1}}{ 1 - \left(p^{\text{up}}_{\max|K} \right)^q } +  \frac{\omega M_K}{1 - p^{\text{mul}}_{\max|K}}, \\
		\nonumber \overbar{T^{\text{DL}}_{\min|K}} &= \omega \max_k n_k \sum_{q=1}^K {K \choose q} \frac{\left( -1 \right)^{q+1}}{ 1 - \left(p^{\text{dist}}_{\min|K} \right)^q } + M_K \frac{\max_k \left\lbrace c_k n_k \right\rbrace }{\epsilon_l } \\
		&+ \omega M_K \sum_{q=1}^K {K \choose q}  \frac{\left( -1 \right)^{q+1}}{ 1 - \left(p^{\text{up}}_{\min|K} \right)^q } +  \frac{\omega M_K}{1 - p^{\text{mul}}_{\min|K}}.		
	\end{align}
\end{proposition}
\begin{IEEEproof}
See Appendix \ref{sec:proof_prop_bound}.
\end{IEEEproof}
From \eqref{ineq:relation_t_dl}, the tightness of the bounds is less than the difference between upper and lower bound. Moreover, the difference between upper and lower bound can be written as
\begin{align}
\nonumber & \overbar{T^{\text{DL}}_{\max|K}} - \overbar{T^{\text{DL}}_{\min|K}}  \\
\nonumber &= \omega \frac{N}{K} \sum_{q=1}^K {K \choose q} \left(-1\right)^{q+1} \left( \frac{1}{1 - \left(p^{\text{dist}}_{\max|K} \right)^q} - \frac{1}{1 - \left(p^{\text{dist}}_{\min|K} \right)^q} \right) \\
\nonumber &+ \omega M_K \sum_{q=1}^K {K \choose q} \left(-1\right)^{q+1} \left( \frac{1}{1 - \left(p^{\text{up}}_{\max|K} \right)^q} - \frac{1}{1 - \left(p^{\text{up}}_{\min|K} \right)^q} \right) \\
& + \omega M_K \left( \frac{1}{1 - p^{\text{mul}}_{\max|K}} - \frac{1}{1 - p^{\text{mul}}_{\min|K}} \right).
\end{align}
Using the following inequality, for $a>0$, $b>0$, and $a>b$,
\begin{align}
\frac{1}{1 - a^q} - \frac{1}{1-b^q} \leq \frac{1}{1-a} - \frac{1}{1-b},
\end{align} we can further bound the gap as
\begin{align}
\nonumber \overbar{T^{\text{DL}}_{\max|K}} - \overbar{T^{\text{DL}}_{\min|K}} & \leq \omega \frac{N}{K} \left(\frac{1}{1 - p^{\text{dist}}_{\max|K}} - \frac{1}{1 - p^{\text{dist}}_{\min|K}} \right) \\
\nonumber &+ \omega M_K \left(\frac{1}{1 - p^{\text{up}}_{\max|K}} - \frac{1}{1 - p^{\text{up}}_{\min|K}} \right) \\
& + \omega M_K \left(\frac{1}{1 - p^{\text{mul}}_{\max|K}} - \frac{1}{1 - p^{\text{mul}}_{\min|K}} \right).
\end{align}
For Rayleigh fading case, we have
\begin{align}
\overbar{T^{\text{DL}}_{\max|K}} - \overbar{T^{\text{DL}}_{\min|K}} & \leq \mathcal{O}\left(M_K\left(\exp\left(\frac{K}{\overbar{\rho^{\min}}} 2^K \right) - \exp\left( \frac{K}{\overbar{\rho^{\max}}}2^K \right) \right) \right).
\end{align}
As the lower and upper bound are derived under the assumption of identical outage probability, if outage probability of edge devices are not highly deviated, the bound can be tight.

Based on Proposition \ref{prop:bounds}, we can find a necessary condition for admitting an additional edge device to participate in distributed learning.
\begin{proposition}\label{prop:addition_device}
If $\overbar{T^{\text{DL}}_{\max|K+1}} - \overbar{T^{\text{DL}}_{\min|K}} \leq 0$, adding an edge device to the distributed edge learning system decreases the average completion time. On the other hand, if $\overbar{T^{\text{DL}}_{\min|K+1}} - \overbar{T^{\text{DL}}_{\max|K}} \geq 0$, the average completion time increases by adding an edge device to the distributed edge learning system.
\end{proposition}
\begin{IEEEproof}
See Appendix \ref{sec:proof_prop_addition}.	
\end{IEEEproof}

Although $\overbar{T^{\text{DL}}_K}$ is not tractable, comparing $\overbar{T^{\text{DL}}_{\min|K}}$ and $\overbar{T^{\text{DL}}_{\max|K}}$, which are expressed in closed form, we can characterize the effect of an additional edge device to the distributed edge learning system. 

\begin{lemma}\label{lem:bound_numtrans}
For $0 \leq p \leq 1$,
	\begin{align}
		\frac{1}{1 - p} \leq \sum_{q=1}^{K} {K\choose q} \frac{(-1)^{q+1}}{1 - p^q} \leq \frac{K}{1 - p}
	\end{align}
\end{lemma}
\begin{IEEEproof}
See Appendix \ref{sec:proof_lem_numtrans}.
\end{IEEEproof}

Lemma \ref{lem:bound_numtrans} provides lower and upper bounds on the number of transmissions that influences the completion time. The lower bound is derived applying Jensen's inequality on max operation. Thus, the lower bound is equivalent to considering only the edge device that requires the maximum number of transmissions. Hence, an additional edge device results in increasing the maximum outage probability, which implies an increase in the maximum number of transmission among edge devices. For the upper bound, the union bound is used. Thus, adding more edge devices increases the total number of edge devices. The effect of increasing the total number of edge devices can be seen in the expression of the upper bound in Lemma \ref{lem:bound_numtrans}.

The necessary condition for allowing an edge device to participate in the distributed learning can be further simplified in the asymptotic regime where $\epsilon_G$ is small.

\begin{proposition}\label{prop:high_precision}
	For $n_k = \frac{N}{K}$ and $c_k = c$, in asymptotic regime where $\epsilon_G \to 0$, if
	\begin{align}
		\nonumber &\exp \left( \frac{1}{\overbar{\eta}^{\max}} \left( 2^{\frac{(K+1)R^{\text{up}}}{B}} -1 \right) \right) + \exp \left( \frac{K+1}{\overbar{\rho}^{\max}} \left( 2^{\frac{R^{\text{mul}}}{B}} - 1 \right) \right) \\
		\nonumber &- K \exp \left( \frac{1}{\overbar{\eta}^{\min}} \left( 2^{\frac{KR^{\text{up}}}{B}} -1 \right) \right) - \exp \left( \frac{K}{\overbar{\rho}^{\min}} \left( 2^{\frac{R^{\text{mul}}}{B}} - 1 \right) \right) \\
		& \geq  \frac{cN}{\epsilon_l K(K+1)}, \label{ineq:prop2}
	\end{align}
	the average completion time increases when an edge device is added to distributed edge learning system.
\end{proposition}
\begin{IEEEproof}
	See Appendix \ref{sec:proof_prop_highprecision}.		
	\end{IEEEproof}

The left-hand side (LHS) of \eqref{ineq:prop2} corresponds to the difference of communication time between the best case of a $K+1$-edge device learning system and the worst case of a $K$-edge device learning system at each global iteration. The right-hand side (RHS) of \eqref{ineq:prop2} can be interpreted as the decrement of local computing time by adding an edge device to the distributed edge learning system. Hence, if an increase in communication time is larger than a decrease in local computing time, the overall average completion time becomes longer.

Meanwhile, when the entire dataset is very large, the time for data distribution and local computing dominates the time taken by other procedures. In other words, the time for local update delivery and global model delivery can be neglected when $N \to \infty$. Therefore, for $N \gg 1$, we can approximate the average completion time as
\begin{align}
	T^{\text{DL}}_K \simeq w \frac{N}{K} \mathbb{E} \left[ \max_{k \in \mathcal{K}} L^{\text{dist}}_{k|K} \right] + M_K \frac{\max_k \left\lbrace c_k n_k \right\rbrace}{\epsilon_l }.
\end{align}
Moreover, if we use the upper bound from Lemma \ref{lem:bound_numtrans}, we can obtain the following upper bound for the average completion time in the large dataset regime
\begin{align}
T^{\text{DL}}_K \leq  \frac{wN}{K} \frac{K}{1 - p^{\text{dist}}_{\max|K}} + M_K \frac{\max_k \left\lbrace c_k n_k \right\rbrace}{\epsilon_l }. \label{ineq:ub_largedata}
\end{align}
For small $K$, we can use the upper bound \eqref{ineq:ub_largedata} as an approximate of the average completion time in the large dataset regime. From Lemma \ref{lem:bound_numtrans}, the performance gap between the upper bound \eqref{ineq:ub_largedata} and the original problem is bounded by the difference between upper bound and lower bound of Lemma \ref{lem:bound_numtrans}.
After some manipulations, we obtain
\begin{align}
T^{\text{DL}+}_K - T^{\text{DL}}_K & \leq \frac{wN}{K} \left( \frac{K-1}{1 - p^{\text{dist}}_{\max|K}} \right),
\end{align} where
\begin{align}
T^{\text{DL}+}_K = \frac{wN}{K} \frac{K}{1 - p^{\text{dist}}_{\max|K}} + M_K\frac{\max_k \left\lbrace c_k n_k \right\rbrace}{\epsilon_l }. \label{eq:ub_largedata2}
\end{align}
Thus, for small $K$, we can use \eqref{eq:ub_largedata2} as an approximation of the average completion time in the large dataset regime. In that case, the optimization problem becomes the minimization of the upper bound of the average completion time, i.e.,
\begin{align}
	 & \min_{K} T^{\text{DL}+}_K  \\
	\text{s.t. } & K \in \mathbb{Z}^+ .
\end{align} 
Relaxing the integer constraint on $K$ for $n_k = \frac{N}{K}$ and $c_k = c$, we can differentiate $T^{\text{DL}+}_K$ as 
\begin{align}
	\nonumber &\frac{\rm d T^{\text{DL}+}_K}{\rm d K} = wN \frac{R^{\text{dist}} \ln 2 }{B\overbar{\rho}^{\min}} 2^{\frac{KR^{\text{dist}}}{B}} \exp \left( \frac{1}{\overbar{\rho}^{\min}} \left( 2^{\frac{KR^{\text{dist}}}{B}} - 1 \right) \right) \\
	& - \frac{wcN}{\left(1 - \epsilon_l \right) \epsilon_l \lambda } \frac{1}{K^2} \ln \left( \frac{\lambda K + 1}{\left(1 - \epsilon_l \right) \epsilon_G \lambda} \right) +  \frac{wcN}{\left(1 - \epsilon_l \right) \epsilon_l} \frac{1}{K}.
\end{align}
Setting the derivative of the objective function to zero, the optimal number of edge devices that minimizes the upper bound of the average completion time in the large dataset regime is the solution of the following equation.
\begin{align}
	 \nonumber &wN \frac{R^{\text{dist}} \ln 2 }{B\overbar{\rho}^{\min}} 2^{\frac{KR^{\text{dist}}}{B}} \exp \left( \frac{1}{\overbar{\rho}^{\min}} \left( 2^{\frac{KR^{\text{dist}}}{B}} - 1 \right) \right)  \\
	 &- \frac{wcN}{\left(1 - \epsilon_l \right) \epsilon_l \lambda } \frac{1}{K^2} \ln \left( \frac{\lambda K + 1}{\left(1 - \epsilon_l \right) \epsilon_G \lambda} \right) +  \frac{wcN}{\left(1 - \epsilon_l \right) \epsilon_l} \frac{1}{K} = 0. \label{eq:optimality_cond}
\end{align}

Using \eqref{eq:optimality_cond}, we can obtain the following results.
\begin{proposition}\label{prop:nec_ub}
	For $n_k = \frac{N}{K}$ and $c_k = c$, $\forall k$, given $\overbar{\rho}^{\min}$, the optimal solution for minimizing the upper bound of the average completion time should satisfy \eqref{eq:nec_cond_ub} in the large dataset regime.
\end{proposition}
\begin{figure*}[t!]
	\begin{align}
	\frac{1}{\overbar{\rho}^{\min}} \geq 2^{-\frac{KR^{\text{dist}}}{B}} \ln \left( \frac{cB}{\epsilon_l \left(1 - \epsilon_l \right) R^{\text{dist}} \ln 2} 2^{-\frac{K R^{\text{dist}}}{B}} \frac{1}{K} \left( \frac{1}{\lambda K} \ln \left(\frac{\lambda K + 1 }{\lambda \left( 1 - \epsilon_l \right) \epsilon_G} \right) - 1 \right) \right) \eqdef Q(K). \label{eq:nec_cond_ub}
	\end{align}
	\hrulefill
\end{figure*}
\begin{IEEEproof}
	See Appendix \ref{sec:proof_prop_necub}.
	\end{IEEEproof}

Proposition \ref{prop:nec_ub} provides a relationship between the average received SNR and the number of edge devices for achieving the minimum upper bound for the average completion time. It can be easily shown that $Q(K)$ is strictly decreasing function of $K$ for $\lambda \leq 1$, $\epsilon_l \leq 1$, and $\epsilon_G \leq 1$. Therefore, as $K$ increases, it is easy to satisfy the necessary condition given in the Proposition \ref{prop:nec_ub}. Also, the LHS of \eqref{eq:nec_cond_ub} is the inverse of the minimum average received SNR. Consequently, both RHS and LHS are decreasing functions of the number of edge devices and the minimum average received SNR, respectively. Thus, if the minimum average received SNR is large, the LHS of \eqref{eq:nec_cond_ub} becomes small. Thus, the optimal $K$ should increase to satisfy inequality \eqref{eq:nec_cond_ub}. In other words, increasing the number of edge devices is necessary when the minimum average received SNR is large. In fact, large minimum average received SNR implies that the distributed edge learning system has high communication capability, which allows to exchange data reliably with less retransmissions. Therefore, allowing more edge devices to participate in the learning process can reduce the average completion time by enhancing parallel computing and sharing the computational load.

However, RHS of \eqref{eq:nec_cond_ub} decreases as inverse exponential of $K$ whereas LHS of \eqref{eq:nec_cond_ub} is inverse of linear function of $\overbar{\rho}^{\min}$. Hence, the speed of decreasing is much faster for RHS than for LHS in \eqref{eq:nec_cond_ub}. This confirms that communication is the major bottleneck in distributed edge learning systems. 

\section{Experiments and Numerical Results}\label{sec:simulations}
In this section, we investigate the behavior of the average completion time through experimental and simulation results in various settings. 
The default simulation environments are $\epsilon_l = 0.001$, $\epsilon_G = 0.001$, $\mu = 1$, $\zeta = 1$, $\lambda = 0.01$, $B = 20$ MHz, $R^{\text{dist}} = 5$ Mbit/s, $R^{\text{up}} = 5$ Mbit/s, $R^{\text{mul}} = 5$ Mbit/s, $\omega = 1$ ms, $\overbar{\rho}^{\min} = 10$ [dB], $\overbar{\rho}^{\max} = 20$ [dB], $\overbar{\eta}^{\min} = 10$ [dB], and $\overbar{\eta}^{\max} = 20$ [dB]. We also assume that the average received SNR of edge devices and the PS, $\overbar{\rho}_k$ and $\overbar{\eta}_k$, are equally spaced in a given interval of $[\overbar{\rho}^{\min}, \overbar{\rho}^{\max}]$ and $[\overbar{\eta}^{\min}, \overbar{\eta}^{\max}]$, respectively. Moreover, we assume that $c_k$ are equally spaced in $[10^{-10}, 10^{-9}]$. For dataset, we use data of SPAM e-mail from \cite{SPAM_dataset} which consists of $4600$ e-mails with $56$ features. We both consider uniform data distribution and non-uniform data distribution.


\begin{figure}[t!]
	\centering
	\includegraphics[scale=0.6]{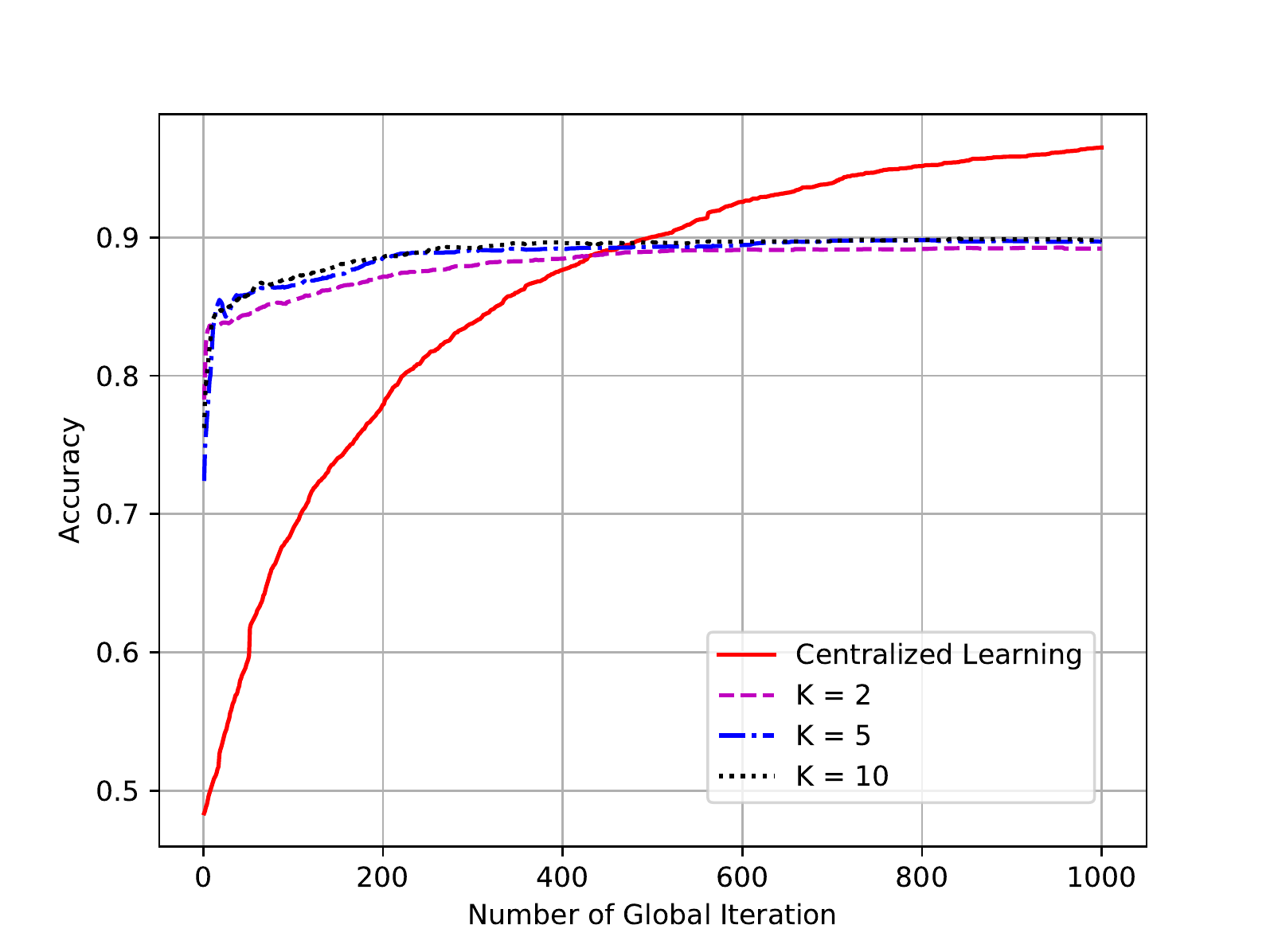}
	\caption{Accuracy of SPAM email detection with Algorithm \ref{alg:cocoa} for different number of edge devices.} \label{fig:exp}
	\end{figure}

First, in Figure \ref{fig:exp}, we verify the convergence of Algorithm \ref{alg:cocoa} through experiments using the real e-mail dataset \cite{SPAM_dataset} and conducting SPAM detection using logistic classification. We can see that accuracy increases rapidly as the number of global iterations increases. Moreover, distributed learning shows accuracy comparable to that of centralized learning.

\begin{figure}[t!]
	\centering
	\includegraphics[scale=0.6]{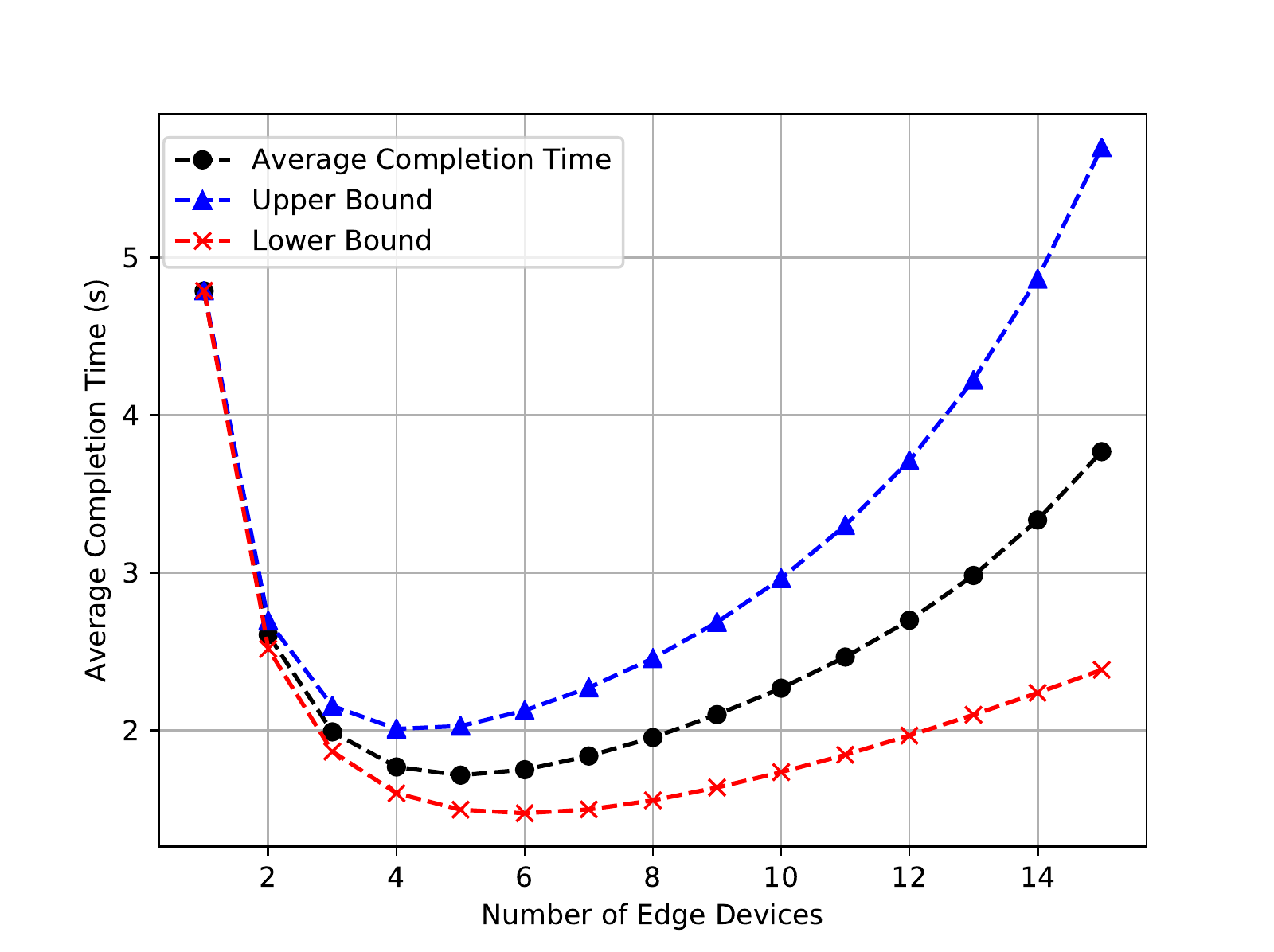}
	\caption{Average completion time and bound for different number of edge devices for uniform data distribution}\label{fig:sim1}
\end{figure}

Under the uniform data distribution, the average completion time, together with its lower and upper bounds are shown in Figure \ref{fig:sim1} when $\overbar{\rho}^{\min} = 10$ [dB], $\overbar{\rho}^{\max} = 20$ [dB], $\overbar{\eta}^{\min} = 10$ [dB], and $\overbar{\eta}^{\max} = 20$ [dB]. Clearly, the average completion time decreases rapidly when the number of edge devices is low, whereas it start increasing past a certain number of edge devices. Intuitively, when the number of edge devices that participates in distributed learning is low, the outage probability is small due to sufficient wireless resources available per edge device. Hence, the average completion time is reduced since the computation time decreases aided by parallel computing as the number of edge devices increases. However, for increasing number of edge devices participating in distributed learning, wireless resources (bandwidth) becomes scarce. Thus, retransmissions are inevitable to deliver data examples, local updates, and global model due to highly likely outages. Consequently, the average completion time starts growing due to longer communication time. Hence, the optimal number of edge devices is the one that balances parallel computing gains and losses from increased communication time. We also observe that the difference between bounds and the average completion time is increasing as the number of edge devices becomes larger. This is expected as for large number of edge devices, more edge devices are treated as having minimum or maximum average received SNR from their true value, thus increasing the gap from the real average completion time.

\begin{figure}[t!]
	\centering
	\includegraphics[scale=0.6]{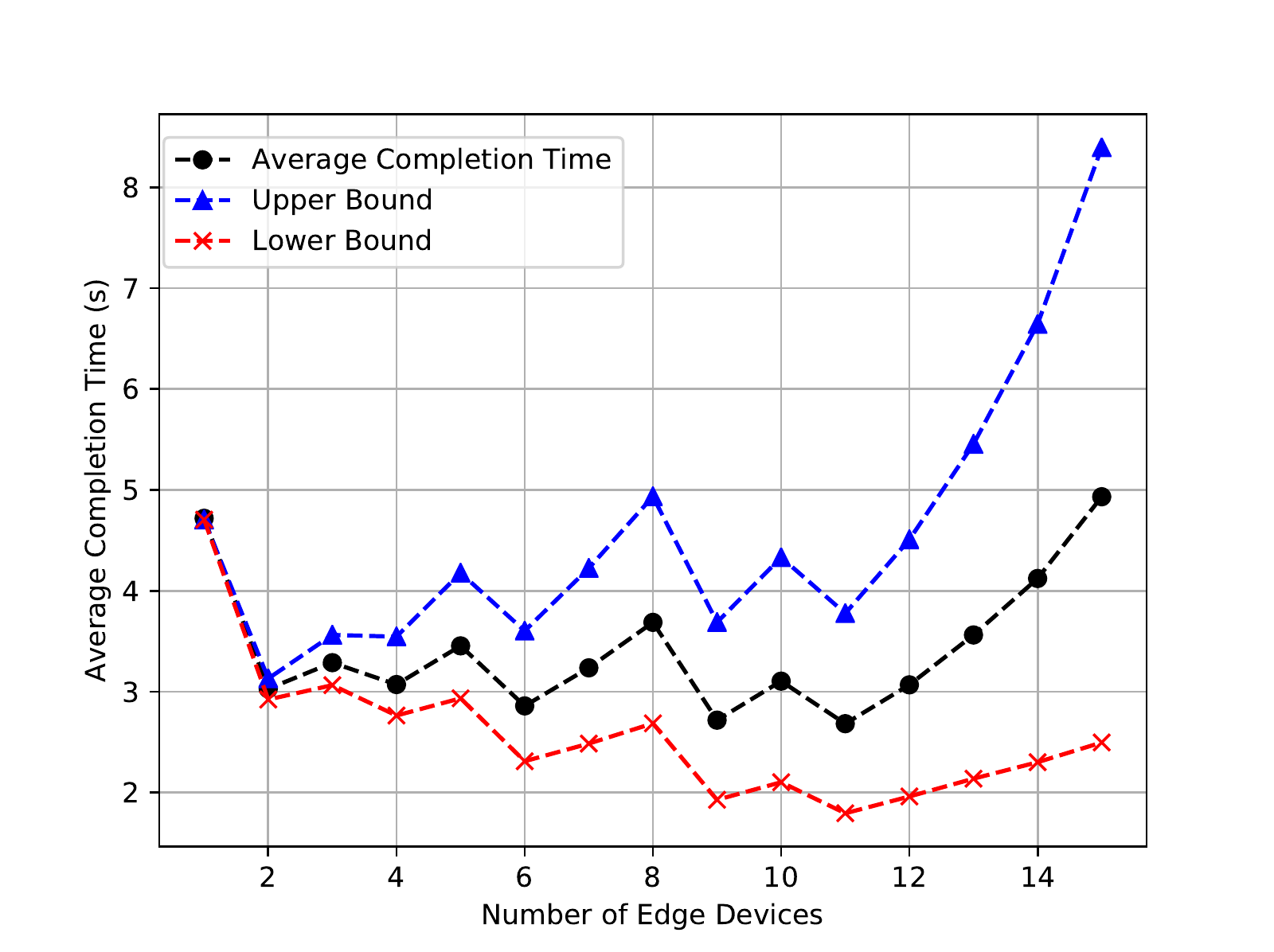}
	\caption{Average completion time and bounds for different number of edge devices for non-uniform data distribution} \label{fig:sim2}
\end{figure}

Figure \ref{fig:sim2} shows the average completion time for non-uniform data distribution. The entire dataset is partitioned into random subsets and given to each edge device. Compared to Figure \ref{fig:sim1}, the average completion time is fluctuating. Since it is possible that a certain edge device receives most of the data for non-uniform data distribution, depending on the realization of data partition, the average completion time changes. However, we can see that the average completion time decreases due to parallel computing for small number of edge devices and increases due to wireless communication overhead for large number of edge devices.

\begin{figure}
	\centering
	\includegraphics[scale=0.6]{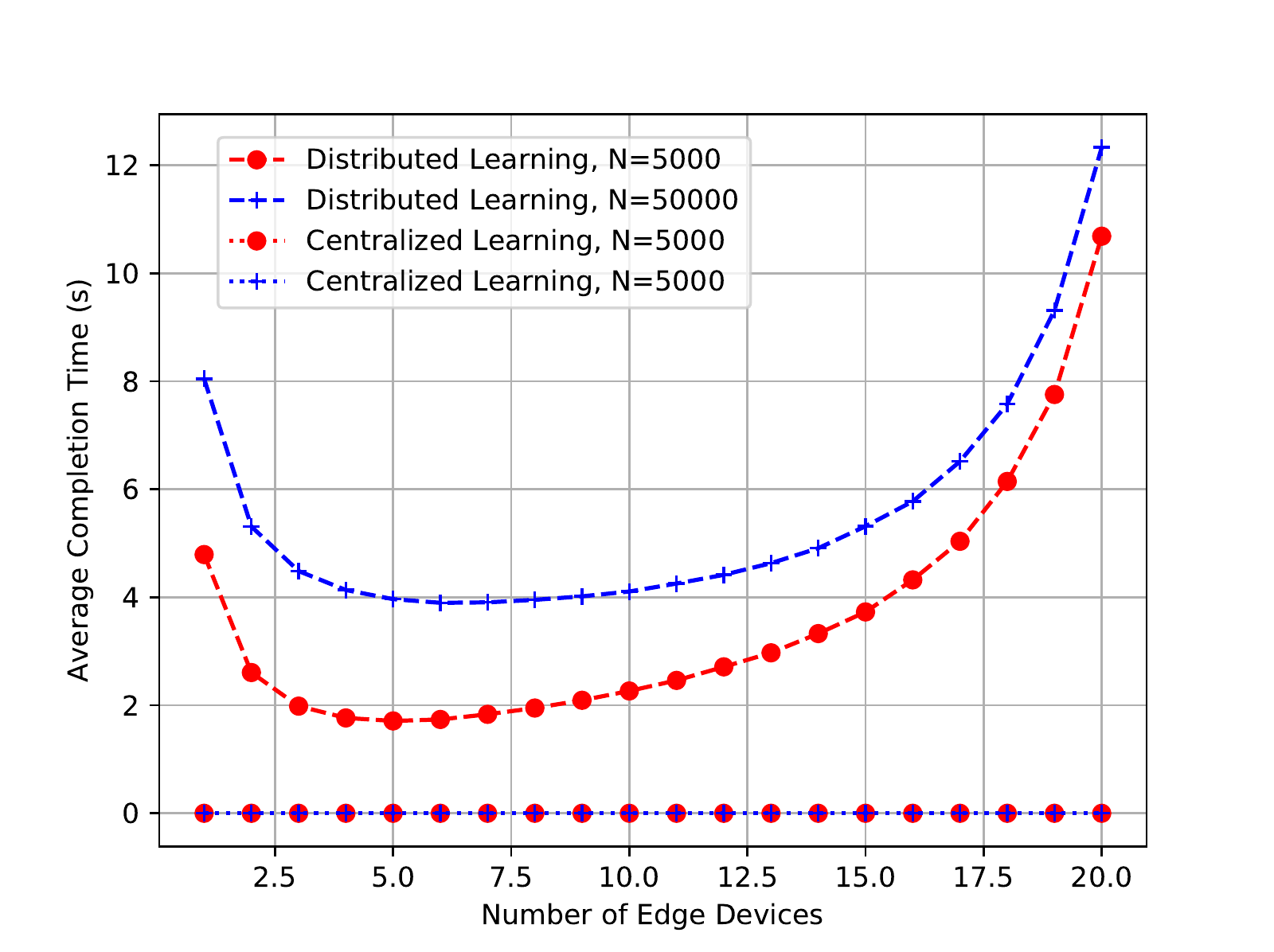}
	\caption{Comparison of the average completion time between centralized learning and distributed learning for different $N$}\label{fig:comp_cl}
\end{figure}

Furthermore, we compare the average completion time of distributed learning with centralized learning in Figure \ref{fig:comp_cl}. By assuming that there exists a central entity that can process based on the whole data, we calculate the average completion time of centralized learning as $T^{\text{central}} = \frac{cN}{\epsilon_G}$. As centralized learning do not need to exchange over the wireless channel and its computing capability is superior to edge devices, centralized learning shows strictly better performance than distributed learning. However, as the number of data increases, the performance gap is reduced due to gain of parallel computation.

\begin{figure}[t!]
	\centering
	\includegraphics[scale=0.6]{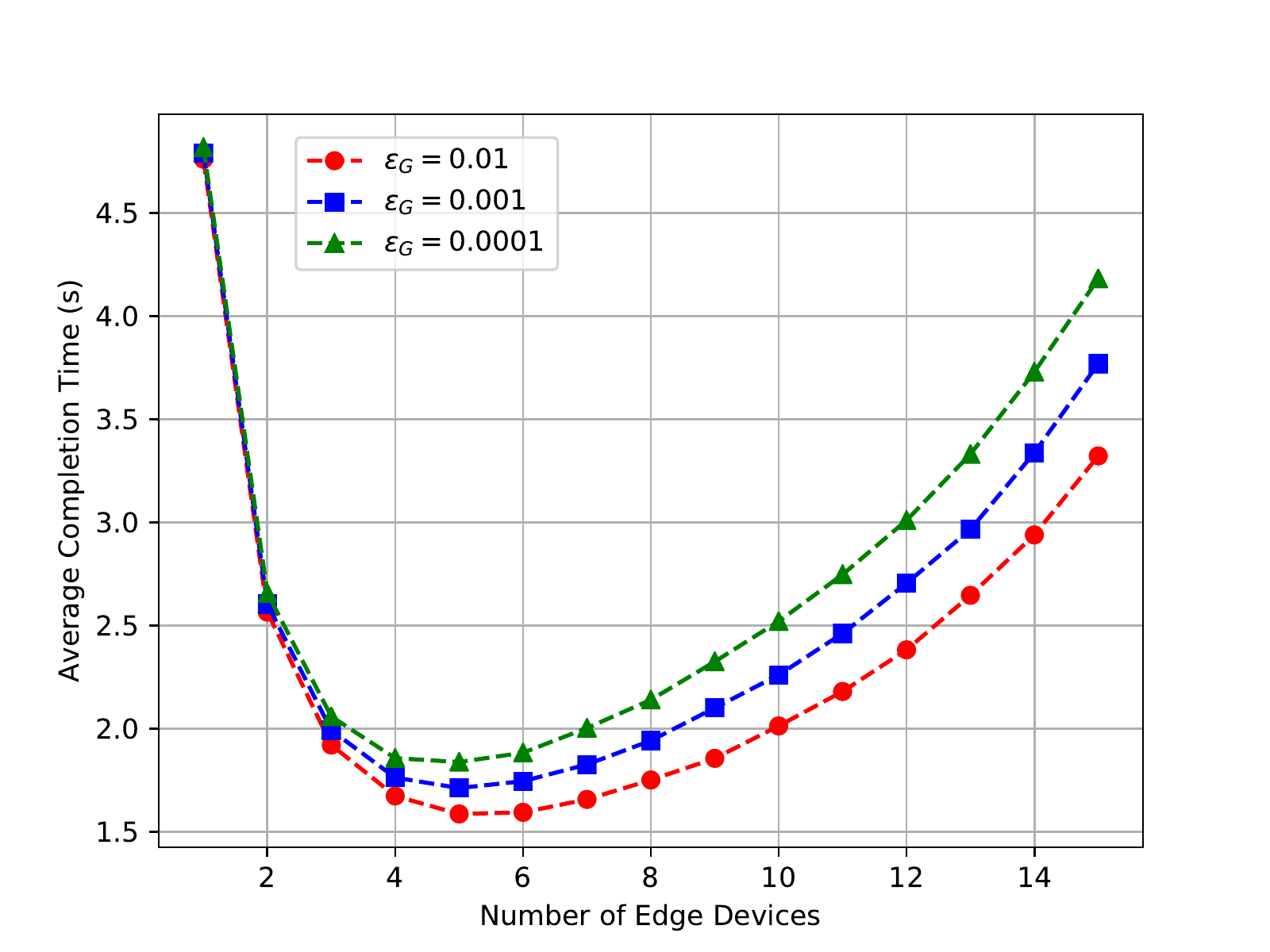}
	\caption{Average completion time and bounds for different number of edge devices when $\overbar{\rho}^{\min} = 10$ [dB], $\overbar{\rho}^{\max} = 20$ [dB], $\overbar{\eta}^{\min} = 10$ [dB], and $\overbar{\eta}^{\max} = 20$ [dB].} \label{fig:sim4}
\end{figure}

The average completion time for different duality gap is shown in the Figure \ref{fig:sim4}. We can see that as the duality gap becomes smaller, the average completion time increases for the entire range of number of edge devices. Since duality gap affects the number of global iterations required, the duality gap does not significantly influence the computation-communication tradeoff. In other words, the more accurate model we aim to obtain, the more communication and computation loads are required. Furthermore, the effect of increasing duality gap to the average completion time is represented in logarithmic scale. Thus, the optimal number of edge devices does not remarkably vary with changing the accuracy of our global model.

\begin{figure}[t!]
	\centering
	\includegraphics[scale=0.6]{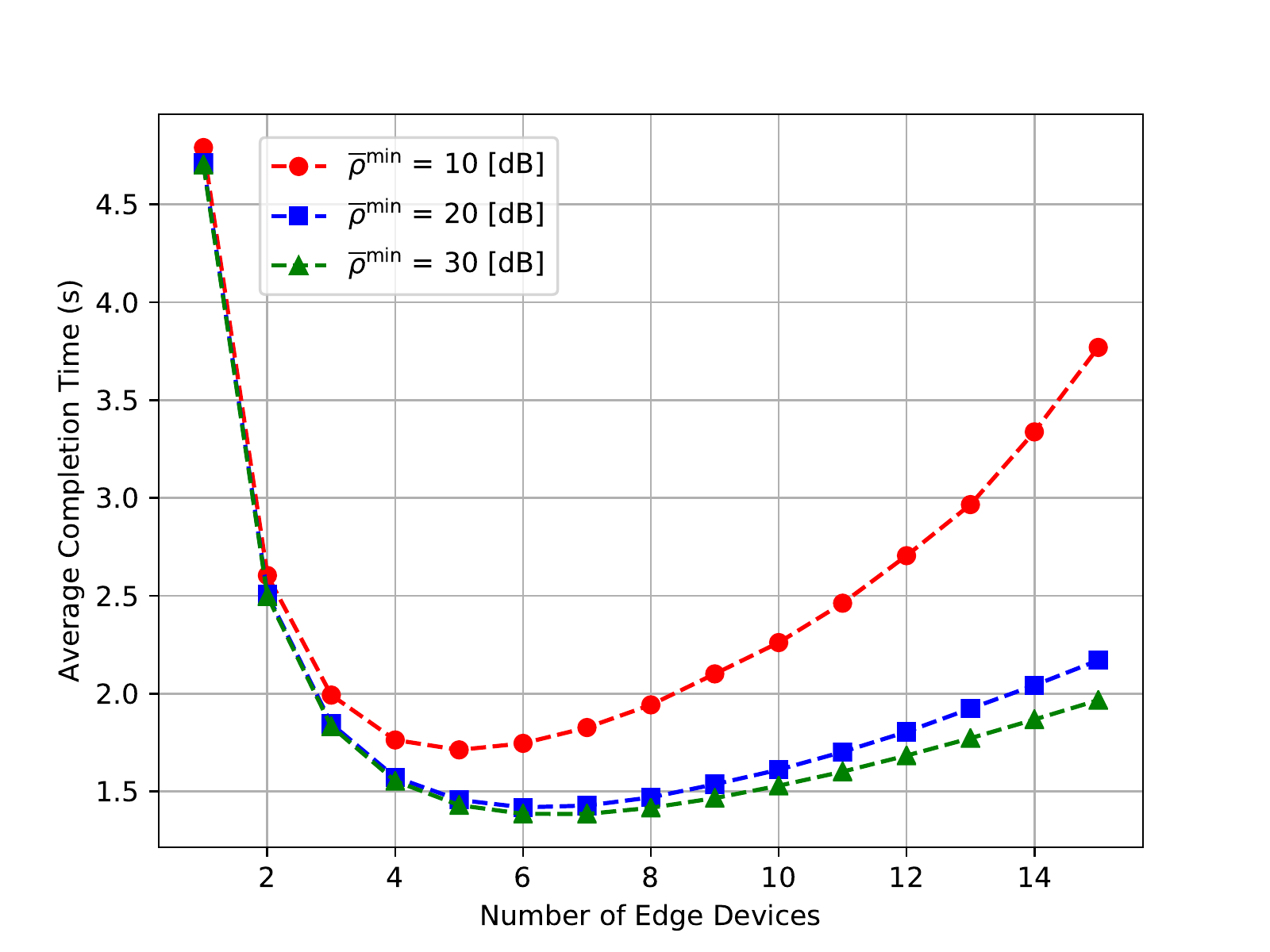}
	\caption{Average completion time for different number of edge devices when $\overbar{\rho}^{\max} =\overbar{\eta}^{\max} = 40$ [dB].} \label{fig:sim7}
\end{figure}

Figure \ref{fig:sim7} shows the effect of the minimum average received SNR to the average completion time. When distributed edge learning algorithm is performed with a large number of edge devices, the minimum average received SNR becomes the critical performance factor. When the minimum average received SNR is low, the average completion time changes dramatically for varying number of edge devices. This implies that when the wireless link is not very stable (or of sufficient quality), the number of edge devices should be carefully configured to guarantee low latency.



\begin{figure}[t!]
	\centering
	\includegraphics[scale=0.6]{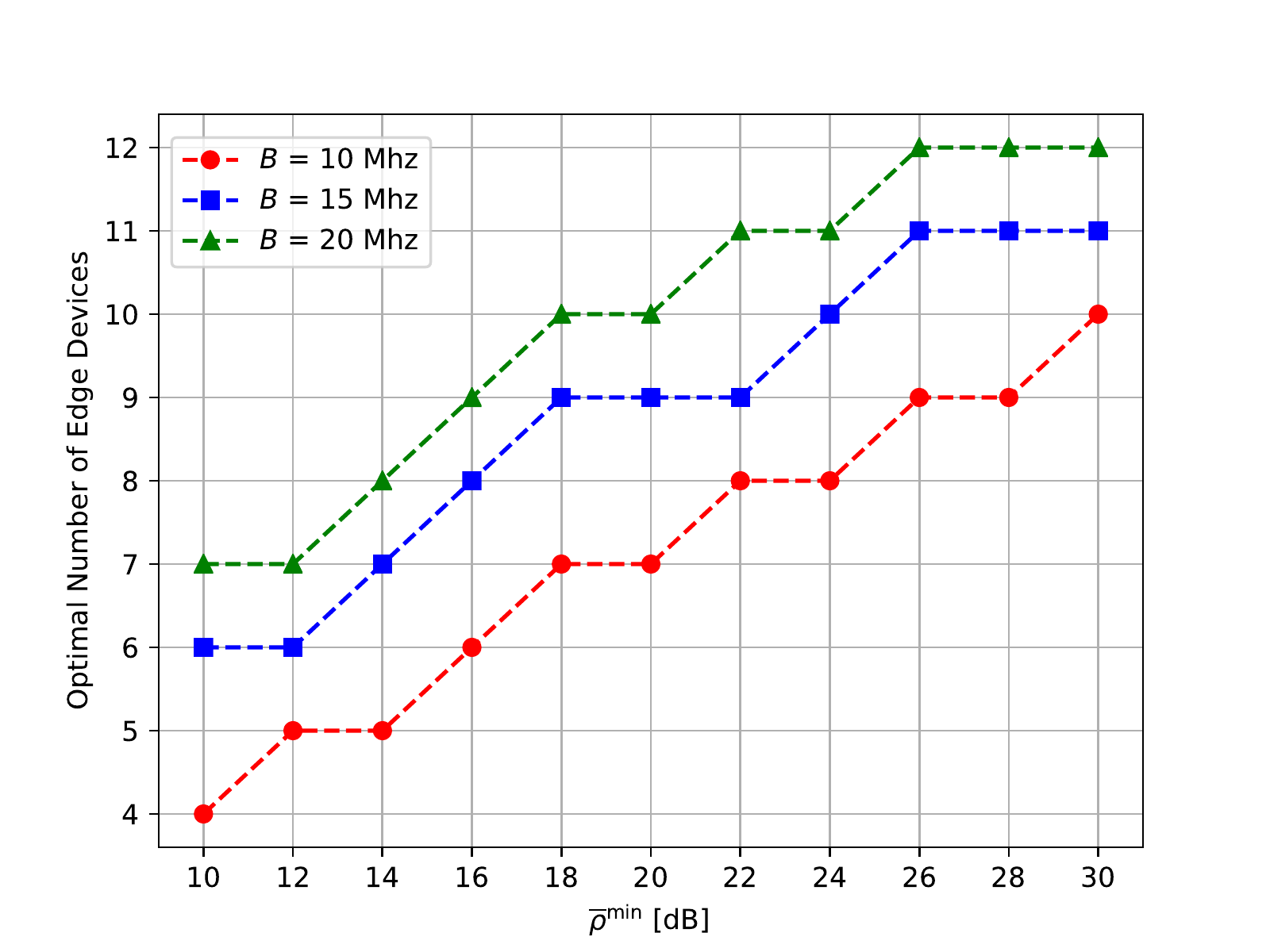}
	\caption{Optimal number of edge devices for different SNRs.} \label{fig:sim8}
\end{figure}

In Figure \ref{fig:sim8}, we plot the optimal number of edge devices as a function of the minimum average received SNR for different bandwidth. Intuitively, as the minimum average received SNR increases, the optimal number of edge devices increases. When the distributed edge learning system can communicate using wider bandwidth, a larger number of edge devices can achieve the minimum average completion time for distributed learning. However, the behavior of the optimal number of edge devices is similar for different bandwidth as the minimum received SNR becomes larger.

\begin{figure}[t!]
	\centering
	\includegraphics[scale=0.6]{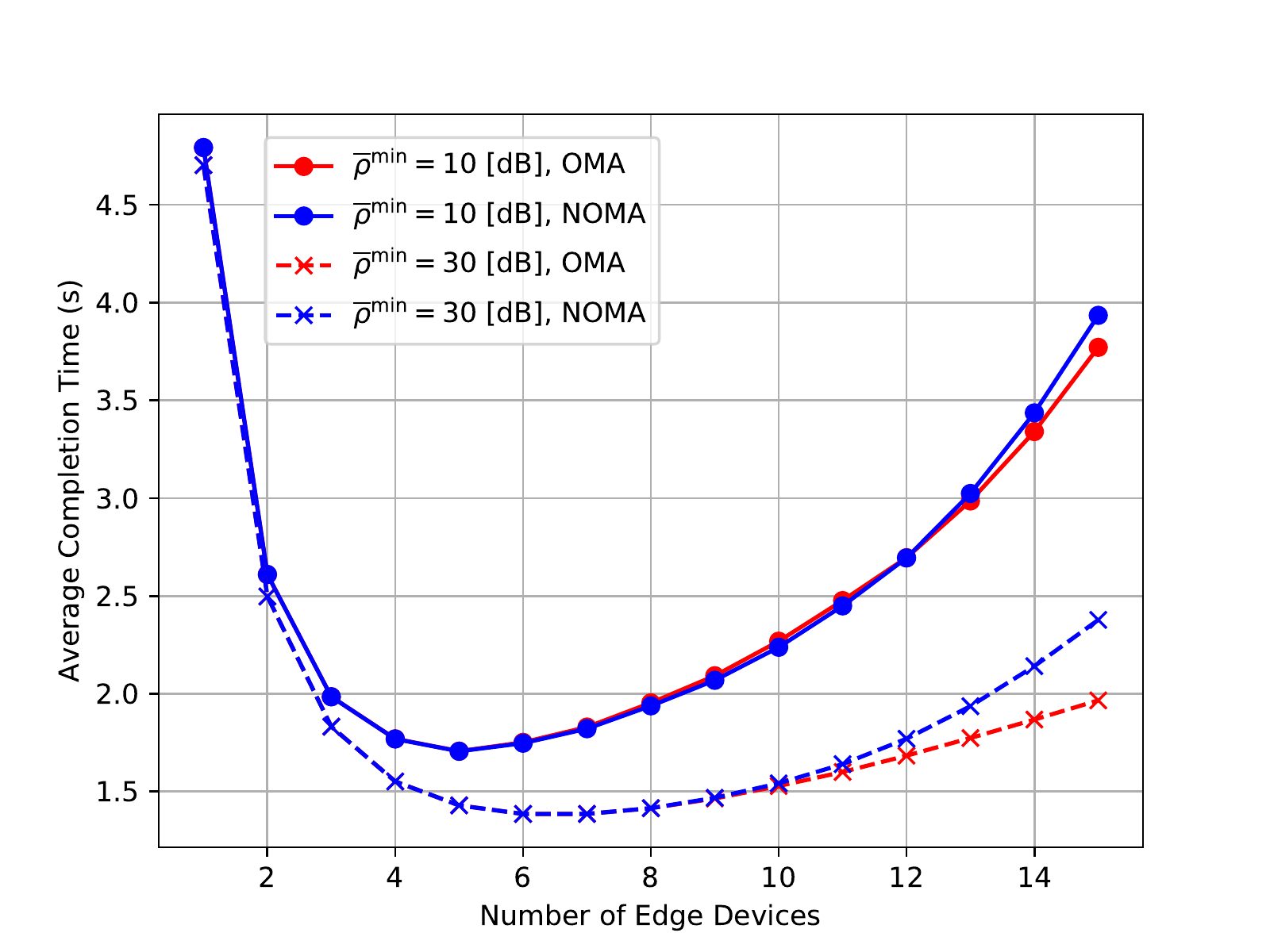}
	\caption{Average completion time of OMA and NOMA for different minimum average received SNRs.} \label{fig:sim10}
\end{figure}

The effect of the multiple access strategy on the performance of distributed edge learning system is shown in Figure \ref{fig:sim10}. Since the communication time for local update delivery depends on the multiple access scheme, we compare the performance of OMA and NOMA. In the NOMA case, we assume that the PS utilizes successive interference cancellation (SIC) and the decoding order is given as the descending order of the received signal strength. Given that the indexes of edge devices are sorted according to descending order of received SNR, the capacity of edge device $k$ for NOMA in local update delivery is given by
\begin{align}
	C_{k|K}^{\text{NOMA}} = B \log \left( 1 + \frac{g_k  P_t^{\text{device}}}{\sum_{j > k} g_j P_t^{\text{device}} + BN_0} \right). \label{eq:capacity_noma}
\end{align}
The outage probability in local update delivery for NOMA is given by
\begin{align}
p^{\text{up, NOMA}}_{k|K} = \mathbb{P} \left[ R^{\text{up}} > B \log \left( 1 + \frac{\eta_k}{\sum_{j > k} \eta_j + 1} \right) \right].
\end{align}

When $\overbar{\rho}^{\min} = 10$ dB, NOMA can achieve lower average completion time than OMA, whereas the overall performance of OMA is superior to that of NOMA for $\overbar{\rho}^{\min} = 30$ dB. Since the achievable rate is highly dependent on bandwidth, NOMA, which utilizes the whole bandwidth, can provide relatively higher rate communication compared with OMA, which allocates partial bandwidth to each edge device. Moreover, in the low SNR regime, interference from other edge devices is relatively weak, making orthogonal bandwidth allocation unnecessary. However, when the received SNR strength is high, orthogonal bandwidth allocation is required to avoid strong interference. 
The minimum average completion time of OMA and NOMA for $\overbar{\rho}^{\min} = 30$ dB is similar and is achieved with the same number of edge devices.

\section{Conclusion}\label{sec:conclusions}

We have studied distributed machine learning at the wireless edge, where edge devices aim to collaboratively train a global model with the help of a remote PS. 
Edge devices have locally distinct partitions of the entire dataset and communicate with the PS over orthogonal wireless fading channels. 
Packets conveying global model and its updates are retransmitted until being successfully received in case of outage events. We have quantified the number of edge devices for which the average completion time is minimized subject to a predefined model accuracy. We have derived upper and lower bounds for the completion time, as well as necessary conditions on the wireless link quality for optimality in the very large dataset and in the high accuracy regimes.  Our analysis confirms that the optimal number achieves balance between communication delay increase and computation time decrease due to parallel processing. As our learning system does not exchange local data except for initial phase, the result of this paper can be used for federated learning. Experimental and simulation results have shown significant completion time speedup when the number of devices participating in the training process is properly set.

\appendices
\section{Proof of Proposition \ref{prop:bounds}}\label{sec:proof_prop_bound}
Let $p^{\text{dist}}_{\max|K}$ represent the maximum outage probability among $K$ edge devices, i.e., $p^{\text{dist}}_{\max|K} = \max_k p^{\text{dist}}_{k|K}$. Considering the worst case where the outage probability of $K$ edge devices is identically given by $p^{\text{dist}}_{\max|K}$, the average completion time of the worst-case scenario should be larger than that of original distributed edge learning system (upper bound). In that case, the number of transmissions to edge devices in data distribution is i.i.d. since edge devices have the same outage probability. If we denote $L_{\max|K}^{\text{dist}}$ as the number of transmissions to an edge device in the worst case, the cumulative distribution function (CDF) of $\max_{k \in \mathcal{K}} \left( L_{\max|K}^{\text{dist}} \right)_k$ can be written as
	\begin{align}
	\mathbb{P} \left[ \max_{k \in \mathcal{K}} \left( L_{\max|K}^{\text{dist}} \right)_k \leq L \right] &= \left( \mathbb{P} \left[ L_{\max|K}^{\text{dist}} \leq L \right] \right)^K.
	\end{align}
	By definition, the PMF of $\max_{k \in \mathcal{K}} \left( L_{\max|K}^{\text{dist}} \right)_k$ becomes
	\begin{align}
	\nonumber &\mathbb{P} \left[ \max_{k \in \mathcal{K}} \left( L_{\max|K}^{+\text{dist}} \right)_k = L \right] 
	\\ & = \left( \mathbb{P} \left[ L_{\max|K}^{\text{dist}} \leq L \right] \right)^K - \left( \mathbb{P} \left[ L_{\max|K}^{\text{dist}} \leq L - 1 \right] \right)^K .
	\end{align}
	
	Using \eqref{eq:PMF_L_dist}, the CDF of $L^{\text{dist}}_{\max|K}$ can be expressed as
	\begin{align}
	\mathbb{P} \left[ L^{\text{dist}}_{\max|K} \leq L \right] &= \sum_{l=1}^L \left(1 - p^{\text{dist}}_{\max|K} \right) \left(p^{\text{dist}}_{\max|K} \right)^{l-1}, \\
	& = 1 - \left(p^{\text{dist}}_{\max|K}\right)^L. \label{eq:CDF_L_k_worst}
	\end{align}  
	Using \eqref{eq:CDF_L_k_worst}, PMF of $\max_{k \in \mathcal{K}} \left( L_{\max|K}^{\text{dist}}\right)_k$ can be rewritten as
	\begin{align}
	\nonumber &\mathbb{P} \left[ \max_{k \in \mathcal{K}} \left( L_{\max|K}^{\text{dist}} \right)_k = L \right] \\
	& = \left( 1 - \left(p^{\text{dist}}_{\max|K} \right)^L \right)^K - \left( 1 - \left(p^{\text{dist}}_{\max|K} \right)^{L-1} \right)^K
	\end{align}
	Consequently, the average number of transmissions in data distribution for the worst-case scenario can be obtained as
	\begin{align}
	\nonumber & \mathbb{E}\left[ \max_{k \in \mathcal{K}} \left( L^{\text{dist}}_{\max|K} \right)_k \right] 
	\\& = \sum_{L=1}^{\infty} L \left[  \left( 1 - \left(p^{\text{dist}}_{\max|K}\right)^L \right)^K - \left( 1 - \left(p^{\text{dist}}_{\max|K}\right)^{L-1} \right)^K \right]. \label{eq:avg_numtx_dist}
	\end{align}
	Using binomial expansion $(a+b)^K = \sum_{q=0}^K {K\choose q} a^{K-q}b^q$, we can expand \eqref{eq:avg_numtx_dist} as
	\begin{align} 
	\nonumber &\mathbb{E}\left[ \max_{k \in \mathcal{K}} \left( L^{\text{dist}}_{\max|K} \right)_k \right] \\
	& = \sum_{L=1}^{\infty} L \left[  \sum_{q=0}^K {K\choose q} \left(-1\right)^q \left(p^{\text{dist}}_{\max|K} \right)^{qL} - \sum_{q=0}^K {K\choose q} \left(-1\right)^q \left(p^{\text{dist}}_{\max|K}\right)^{q(L-1)} \right] , \\
	& = \sum_{q=0}^K {K \choose q}\left( -1 \right)^q \sum_{L=1}^{\infty} L \left[ \left(p^{\text{dist}}_{\max|K} \right)^{qL} - \left(p^{\text{dist}}_{\max|K} \right)^{q\left(L-1\right)} \right].
	\end{align}
	After algebraic manipulations, the average number of transmissions in data distribution for the worst case becomes
	\begin{align}
	\mathbb{E}\left[ \max_{k \in \mathcal{K}} \left( L^{\text{dist}}_{\max|K} \right)_k \right] 
	& = \sum_{q=1}^K {K \choose q}\left( -1 \right)^{q+1}  \frac{1}{ 1 - \left(p^{\text{dist}}_{\max|K} \right)^q } \label{eq:L_dist_worst}.
	\end{align}
	
With similar derivation, we can calculate the number of transmissions in the other distributed learning phases for the worst case. If we denote the number of transmissions in local update delivery for the worst-case scenario as $\max_{k \in \mathcal{K}} \left( L^{\text{up}}_{\max|K} \right)_k$,
	\begin{align}
	\mathbb{E} \left[ \max_{k \in \mathcal{K}}\left( L^{\text{up}}_{\max|K} \right)_k \right] &= \sum_{q=1}^K {K \choose q}\left( -1 \right)^{q+1}  \frac{1}{ 1 - \left(p^{\text{up}}_{\max|K} \right)^q }, \label{eq:L_up_worst}
	\end{align} 
	where $p^{\text{up}}_{\max|K}$ is the maximum outage probability of edge devices in local update delivery.
	
Unlike other procedures, in global model delivery, multicast transmission is utilized. Thus, if the edge device experiencing the minimum channel gain can receive the global model successfully, all edge devices can receive the global model without outage. In this case, the PMF of the number of transmissions in global model delivery is given as
	\begin{align}
	\mathbb{P} \left[ L^{\text{mul}}_{\max|K} = L \right] &= \left(1 - p^{\text{mul}}_{\max|K} \right) \left( p^{\text{mul}}_{\max|K} \right)^{L-1} , \label{eq:PMF_mul_worst}
	\end{align} 
where $p^{\text{mul}}_{\max|K}$ is outage probability in multicast transmission with $K$ edge devices for the worst case.
	
	Using \eqref{eq:PMF_mul_worst}, we can obtain the average number of transmissions in global model delivery.
	\begin{align}
	\mathbb{E} \left[ L^{\text{mul}}_{\max|K} \right] &= \sum_{L=1}^{\infty} L \left(1 - p^{\text{mul}}_{\max|K} \right) \left( p^{\text{mul}}_{\max|K} \right)^{L-1}, \\
	& = \frac{1}{1 - p^{\text{mul}}_{\max|K}} \label{eq:L_mul_worst}
	\end{align}
Based on \eqref{eq:L_dist_worst}, \eqref{eq:L_up_worst}, and \eqref{eq:L_mul_worst}, the completion time for the worst case can be represented as
	\begin{align}
	\nonumber &\overbar{T^{\text{DL}}_{\max|K}} = \omega \max_k \left\lbrace n_k \right\rbrace  \mathbb{E}\left[ \max_{k \in \mathcal{K}} \left( L^{\text{dist}}_{\max|K} \right)_k \right] + M_K \max_{k \in \mathcal{K}} T^{\text{local}}_k \\
	& + M_K \omega \mathbb{E}\left[ \max_{k \in \mathcal{K}} \left( L^{\text{up}}_{\max|K} \right)_k \right] + M_K \omega \mathbb{E}\left[ L^{\text{mul}}_{\max|K} \right], \\
	\nonumber & = \omega \max_k \left\lbrace n_k \right\rbrace \sum_{q=1}^K {K \choose q}  \frac{\left( -1 \right)^{q+1}}{ 1 - \left(p^{\text{dist}}_{\max|K} \right)^q } + M_K \frac{\max_k \left\lbrace c_k n_k \right\rbrace }{\epsilon_l} \\
	&+ \omega M_K \sum_{q=1}^K {K \choose q}  \frac{\left( -1 \right)^{q+1}}{ 1 - \left(p^{\text{up}}_{\max|K} \right)^q } + \omega M_K \frac{1}{1 - p^{\text{mul}}_{\max|K}}.
	\end{align}
	
On the other hand, as a lower bound, we can consider the best case in which the outage probability of all edge devices is the minimum outage probability, thus the average completion time becomes lower than that of original system. If we denote the number of transmissions in data distribution, local updates delivery, and global model delivery as $L^{\text{dist}}_{\min|K}$, $L^{\text{up}}_{\min|K}$ and $L^{\text{mul}}_{\min|K}$ for the best case, respectively, we can represent the number of transmissions for the corresponding procedures as follows.
	\begin{align}
	\mathbb{E}\left[ \max_{k \in \mathcal{K}} \left( L^{\text{dist}}_{\min|K} \right)_k \right] &= \sum_{q=1}^K {K \choose q}\left( -1 \right)^{q+1}  \frac{1}{ 1 - \left(p^{\text{dist}}_{\min|K} \right)^q } ,\\
	\mathbb{E}\left[ \max_{k \in \mathcal{K}} \left( L^{\text{up}}_{\min|K} \right)_k \right] &= \sum_{q=1}^K {K \choose q}\left( -1 \right)^{q+1}  \frac{1}{ 1 - \left(p^{\text{up}}_{\min|K} \right)^q } , \\
	\mathbb{E}\left[ L^{\text{mul}}_{\min|K} \right] & = \frac{1}{1 - p^{\text{mul}}_{\min|K}},
	\end{align} 
	where $p^{\text{dist}}_{\min|K}$, $p^{\text{up}}_{\min|K}$ and $p^{\text{mul}}_{\min|K}$ are the minimum outage probability in data distribution, local update delivery and global model delivery, respectively.
	
Accordingly, the average completion time of the best case can be obtained as
	\begin{align}
	\nonumber & \overbar{T^{\text{DL}}_{\min|K}} = \omega \max_k \left\lbrace n_k \right\rbrace \mathbb{E}\left[ \max_{k \in \mathcal{K}} \left( L^{\text{dist}}_{\min|K} \right)_k \right] + M_K \max_{k \in \mathcal{K}} T^{\text{local}}_k \\
	&+ M_K \omega \mathbb{E}\left[ \max_{k \in \mathcal{K}} \left( L^{\text{up}}_{\min|K} \right)_k \right] + M_K \omega \mathbb{E}\left[ L^{\text{mul}}_{\min|K} \right], \\
	\nonumber & = \omega \max_k \left\lbrace n_k \right\rbrace \sum_{q=1}^K {K \choose q}  \frac{\left( -1 \right)^{q+1}}{ 1 - \left(p^{\text{dist}}_{\min|K} \right)^q } + M_K \frac{\max_k \left\lbrace c_k n_k \right\rbrace }{\epsilon_l} \\
	&+ \omega M_K \sum_{q=1}^K {K \choose q}  \frac{\left( -1 \right)^{q+1}}{ 1 - \left(p^{\text{up}}_{\min|K} \right)^q } + \omega M_K \frac{1}{1 - p^{\text{mul}}_{\min|K}}.
	\end{align}

\section{Proof of Proposition \ref{prop:addition_device}}\label{sec:proof_prop_addition}
	From \eqref{ineq:relation_t_dl}, we have the following inequalities.
	\begin{align}
	\overbar{T^{\text{DL}}_{\min|K}} \leq \overbar{T^{\text{DL}}_K} \leq \overbar{T^{\text{DL}}_{\max|K}}, \label{ineq:prop_temp_1} \\
	-\overbar{T^{\text{DL}}_{\max|K+1}} \leq -\overbar{T^{\text{DL}}_{K+1}} \leq -\overbar{T^{\text{DL}}_{\min|K+1}} \label{ineq:prop_temp_2}.
	\end{align}
	If we combine \eqref{ineq:prop_temp_1} and \eqref{ineq:prop_temp_2}, we have
	\begin{align}
	\overbar{T^{\text{DL}}_{\min|K+1}} - \overbar{T^{\text{DL}}_{\max|K}} \leq  \overbar{T^{\text{DL}}_{K+1}} - \overbar{T^{\text{DL}}_K} \leq \overbar{T^{\text{DL}}_{\max|K+1}} - \overbar{T^{\text{DL}}_{\min|K}}. \label{ineq:delta_t_dl}
	\end{align}
	Hence, if $\overbar{T^{\text{DL}}_{\max|K+1}} - \overbar{T^{\text{DL}}_{\min|K}} \leq 0$, from \eqref{ineq:delta_t_dl},
	\begin{align}
	\overbar{T^{\text{DL}}_{K+1}} - \overbar{T^{\text{DL}}_K} \leq 0.
	\end{align}
	Consequently, $\overbar{T^{\text{DL}}_{K+1}} \leq  \overbar{T^{\text{DL}}_K} $. Thus, the average completion time is reduced after adding an edge device to the distributed edge learning system.
	
	On the other hand, if $\overbar{T^{\text{DL}}_{\min|K+1}} - \overbar{T^{\text{DL}}_{\max|K}} \geq 0$, we have $\overbar{T^{\text{DL}}_{K+1}} - \overbar{T^{\text{DL}}_K} \geq 0$.
	Thus, $\overbar{T^{\text{DL}}_{\min|K+1}} \geq  \overbar{T^{\text{DL}}_{\max|K}}$ implies that the average completion time increases by adding one edge device in the distributed edge learning.
	
\section{Proof of Lemma \ref{lem:bound_numtrans}}\label{sec:proof_lem_numtrans}
Suppose we have $K$ edge devices and that the outage probability when transmitting to them is given by $p$. 
We denote $L_k$ the number of transmissions to the $k$-th edge device. Then, with similar derivation as for \eqref{eq:L_dist_worst}, we have
	\begin{align}
	\mathbb{E}\left[ \max_k L_k \right] = \sum_{q=1}^{K} {K\choose q} \frac{(-1)^{q+1}}{1 - p^q}.
	\end{align}
In fact, the maximum of the average number of transmissions is equal to the average number of transmissions for the largest outage probability. Since the maximum of the number of transmissions among $K$ edge devices is always larger than the number of transmissions for the edge device with the highest outage probability, the maximum of the average number of transmissions should be less than the average of the maximum number of transmissions. Formally, we have  
	\begin{align}
	\max_k \mathbb{E}\left[ L_k \right] \leq  \mathbb{E}\left[ \max_k L_k \right]. \label{ineq:lemma_temp1}
	\end{align}
	The sum of the number of transmissions for edge devices is obviously larger than the maximum of the number of transmissions. Hence,
	\begin{align}
	\mathbb{E}\left[ \max_k L_k \right] \leq \sum_{k=1}^K \mathbb{E}\left[ L_k \right]. \label{ineq:lemma_temp2}
	\end{align}
	Also, the average number of transmissions with outage probability $p$ is given as
	\begin{align}
		\mathbb{E}\left[ L_k \right] = \frac{1}{1- p}. \label{eq:lemma_temp1}
	\end{align}
	Combining \eqref{ineq:lemma_temp1}, \eqref{ineq:lemma_temp2} and \eqref{eq:lemma_temp1}, we have
	\begin{align}
	\frac{1}{1 - p} \leq \sum_{q=1}^{K} {K\choose q} \frac{(-1)^{q+1}}{1 -  p^q} \leq \frac{K}{1 - p}.
	\end{align}

\section{Proof of Proposition \ref{prop:high_precision}}\label{sec:proof_prop_highprecision}
As $\epsilon_G \to 0$, the number of global iterations required increases to infinity. Hence, the average completion time is dominated by the time taken by repeated procedures in every global iteration. In other words, the time for data distribution could be neglected. Hence, for $n_k = \frac{N}{K}$ and $c_k = c$, in the high precision regime (small $\epsilon_G$), we can approximate the upper and lower bound of the average completion time as \eqref{eq:prop_temp1} and \eqref{eq:prop_temp2}, respectively.
	\begin{figure*}[t!]
		\begin{align}
		\overbar{T^{\text{DL}}_{\min|K+1}} & \simeq w M_{K+1} \left( \frac{N}{\epsilon_l (K+1)} + \sum_{q=1}^{K+1} {K+1\choose q} \frac{(-1)^{q+1}}{1 - \left( p^{\text{up}}_{\min|K+1} \right)^q} + \frac{1}{1 - p^{\text{mul}}_{\min|K+1}} \right) \label{eq:prop_temp1} \\
		\overbar{T^{\text{DL}}_{\max|K}} & \simeq w M_K \left( \frac{cN}{\epsilon_l K} + \sum_{q=1}^{K} {K\choose q} \frac{(-1)^{q+1}}{1 - \left( p^{\text{up}}_{\max|K} \right)^q} + \frac{1}{1 - p^{\text{mul}}_{\max|K}} \right)\label{eq:prop_temp2}
		\end{align}
		\hrulefill
	\end{figure*}
	Furthermore, since $M_K$ is an increasing function of $K$, \eqref{eq:prop_temp2} can be bounded above as
	\begin{align}
	\overbar{T^{\text{DL}}_{\max|K}} & \leq w M_{K+1} \left( \frac{cN}{\epsilon_l K} + \sum_{q=1}^{K} {K\choose q} \frac{(-1)^{q+1}}{1 - \left( p^{\text{up}}_{\max|K} \right)^q} + \frac{1}{1 - p^{\text{mul}}_{\max|K}} \right).
	\end{align}
	Using the upper bound from Lemma \ref{lem:bound_numtrans},
	\begin{align}
	\overbar{T^{\text{DL}}_{\max|K}} & \leq w M_{K+1} \left( \frac{cN}{\epsilon_l K} + \frac{K}{1 - p^{\text{up}}_{\max|K}} + \frac{1}{1 - p^{\text{mul}}_{\max|K}} \right) \label{ineq:prop_temp2}
	\end{align}
	If we also use the lower bound from Lemma \ref{lem:bound_numtrans} to \eqref{eq:prop_temp1},
	\begin{align}
	\overbar{T^{\text{DL}}_{\min|K+1}} \geq w M_{K+1} \left( \frac{cN}{\epsilon_l (K+1)} + \frac{1}{1 - p^{\text{up}}_{\min|K+1}} + \frac{1}{1 - p^{\text{mul}}_{\min|K+1}} \right) \label{ineq:prop_temp3}
	\end{align}
		\begin{figure*}[t!]
		\begin{align}	
		\overbar{T^{\text{DL}}_{\min|K+1}} - \overbar{T^{\text{DL}}_{\max|K}} & \geq w M_{K+1} \left( \frac{cN}{\epsilon_l (K+1)} + \frac{1}{1 - p^{\text{up}}_{\min|K+1}} + \frac{1}{1 - p^{\text{mul}}_{\min|K+1}} - \frac{cN}{\epsilon_l K} - \frac{K}{1 - p^{\text{up}}_{\max|K}} - \frac{1}{1 - p^{\text{mul}}_{\max|K}} \right). \label{ineq:prop2_temp1}
		\end{align}
		\hrulefill
	\end{figure*}
	Using \eqref{ineq:prop_temp2} and \eqref{ineq:prop_temp3}, we have \eqref{ineq:prop2_temp1}.
	
Using the CDF for Rayleigh fading and order statistics, the maximum and the minimum outage probability for local updated delivery and global model delivery can be represented respectively by
	\begin{align}
	p^{\text{up}}_{\max|K} &= 1 - \exp \left( -\frac{1}{K\overbar{\eta}^{\min}} \left( 2^{\frac{KR^{\text{up}}}{B}} -1 \right) \right), \label{eq:outage_1} \\
	p^{\text{up}}_{\min|K+1} &= 1 - \exp \left( -\frac{1}{K\overbar{\eta}^{\max}} \left( 2^{\frac{(K+1)R^{\text{up}}}{B}} -1 \right) \right) , \label{eq:outage_2}\\
	p^{\text{mul}}_{\max|K} &= 1 - \exp \left( -\frac{K}{\overbar{\rho}^{\min}} \left( 2^{\frac{R^{\text{mul}}}{B}} - 1 \right) \right), \label{eq:outage_3} \\
	p^{\text{mul}}_{\max|K+1} &= 1 - \exp \left( -\frac{K+1}{\overbar{\rho}^{\max}} \left( 2^{\frac{R^{\text{mul}}}{B}} - 1 \right) \right), \label{eq:outage_4} 
	\end{align} where $\overbar{\eta}^{\min} = \min_k \mathbb{E}\left[ \eta_k \right]$, $\overbar{\eta}^{\max} = \max_k \mathbb{E}\left[ \eta_k \right]$, $\overbar{\rho}^{\min} = \min_k \mathbb{E}\left[ \rho_k \right]$, and $\overbar{\rho}^{\max} = \max_k \mathbb{E}\left[ \rho_k \right]$.
	
After substitution of outage probabilities with \eqref{eq:outage_1}, \eqref{eq:outage_2} \eqref{eq:outage_3}, and \eqref{eq:outage_4} correspondingly, \eqref{ineq:prop2_temp1} can be rewritten as
	\begin{align}
	\nonumber &\overbar{T^{\text{DL}}_{\min|K+1}} - \overbar{T^{\text{DL}}_{\max|K}}\\
	 \nonumber &\geq w M_{K+1} \left[ \frac{cN}{\epsilon_l (K+1)} + \exp \left( \frac{1}{K\overbar{\eta}^{\max}} \left( 2^{\frac{(K+1)R^{\text{up}}}{B}} -1 \right) \right) \right.\\
	 \nonumber &\left. + \exp \left( \frac{K+1}{\overbar{\rho}^{\max}} \left( 2^{\frac{R^{\text{mul}}}{B}} - 1 \right) \right) - \frac{cN}{\epsilon_l K} - K \exp \left( \frac{1}{K\overbar{\eta}^{\min}} \left( 2^{\frac{KR^{\text{up}}}{B}} -1 \right) \right) \right. \\
	 & \left.- \exp \left( \frac{K}{\overbar{\rho}^{\min}} \left( 2^{\frac{R^{\text{mul}}}{B}} - 1 \right) \right) \right] , \\
	\nonumber & = w M_{K+1} \left[ \exp \left( \frac{1}{K\overbar{\eta}^{\max}} \left( 2^{\frac{(K+1)R^{\text{up}}}{B}} -1 \right) \right) \right. \\
	\nonumber & \left. + \exp \left( \frac{K+1}{\overbar{\rho}^{\max}} \left( 2^{\frac{R^{\text{mul}}}{B}} - 1 \right) \right) - K \exp \left( \frac{1}{K\overbar{\eta}^{\min}} \left( 2^{\frac{KR^{\text{up}}}{B}} -1 \right) \right) \right. \\
	&\left. - \exp \left( \frac{K}{\overbar{\rho}^{\min}} \left( 2^{\frac{R^{\text{mul}}}{B}} - 1 \right) \right) - \frac{cN}{\epsilon_l K(K+1)}  \right]. 
	\end{align}
	Since $w M_{K+1}$ is always positive, if the following inequality holds,
	\begin{align}
	\nonumber &\exp \left( \frac{1}{K\overbar{\eta}^{\max}} \left( 2^{\frac{(K+1)R^{\text{up}}}{B}} -1 \right) \right) + \exp \left( \frac{K+1}{\overbar{\rho}^{\max}} \left( 2^{\frac{R^{\text{mul}}}{B}} - 1 \right) \right) \\
	\nonumber &- K \exp \left( \frac{1}{K\overbar{\eta}^{\min}} \left( 2^{\frac{KR^{\text{up}}}{B}} -1 \right) \right) - \exp \left( \frac{K}{\overbar{\rho}^{\min}} \left( 2^{\frac{R^{\text{mul}}}{B}} - 1 \right) \right)  \\
	&\geq  \frac{cN}{\epsilon_l K(K+1)},
	\end{align}
	we have
	\begin{align}
	\overbar{T^{\text{DL}}_{\min|K+1}} - \overbar{T^{\text{DL}}_{\max|K}} \geq 0 . \label{ineq:prop_2}
	\end{align}	
Therefore, by Proposition \ref{prop:addition_device}, \eqref{ineq:prop_2} implies that the average completion time increases when an additional edge device participates in the distributed edge learning.
	
\section{Proof of Proposition \ref{prop:nec_ub}}\label{sec:proof_prop_necub}
If we rewrite \eqref{eq:optimality_cond}, we have
\begin{align}
	\nonumber &wN \frac{R^{\text{dist}} \ln 2 }{B\overbar{\rho}^{\min}} 2^{\frac{KR^{\text{dist}}}{B}} \exp \left( \frac{1}{\overbar{\rho}^{\min}} \left( 2^{\frac{KR^{\text{dist}}}{B}} - 1 \right) \right) \\
	&= \frac{wcN}{\left(1 - \epsilon_l \right) \epsilon_l \lambda } \frac{1}{K^2} \ln \left( \frac{\lambda K + 1}{\left(1 - \epsilon_l \right) \epsilon_G \lambda} \right) - \frac{wcN}{\left(1 - \epsilon_l \right) \epsilon_l} \frac{1}{K}.
	\end{align}
Since $\frac{1}{\overbar{\rho}^{\min}} < \exp \left(\frac{1}{\overbar{\rho}^{\min}}\right)$, 
	\begin{align}
	\nonumber &wN \frac{R^{\text{dist}} \ln 2 }{B \rho^{\min}} 2^{\frac{KR^{\text{dist}}}{B}} \exp \left( \frac{1}{\overbar{\rho}^{\min}} \left( 2^{\frac{KR^{\text{dist}}}{B}} - 1 \right) \right) \\
	&< wN \frac{R^{\text{dist}} \ln 2 }{B} 2^{\frac{KR^{\text{dist}}}{B}} \exp \left( \frac{1}{\overbar{\rho}^{\min}} 2^{\frac{KR^{\text{dist}}}{B}} \right).
	\end{align}
Therefore, if the following inequality holds
	\begin{align}
	\nonumber &wN \frac{R^{\text{dist}} \ln 2 }{B} 2^{\frac{KR^{\text{dist}}}{B}} \exp \left( \frac{1}{\overbar{\rho}^{\min}} 2^{\frac{KR^{\text{dist}}}{B}} \right) \\
	& < \frac{wcN}{\left(1 - \epsilon_l \right) \epsilon_l \lambda } \frac{1}{K^2} \ln \left( \frac{\lambda K + 1}{\left(1 - \epsilon_l \right) \epsilon_G \lambda} \right) - \frac{wcN}{\left(1 - \epsilon_l \right) \epsilon_l} \frac{1}{K} \label{eq:prop3_temp2}
	\end{align}
	then, we have
	\begin{align}
	\nonumber &wN \frac{R^{\text{dist}} \ln 2 }{B\overbar{\rho}^{\min}} 2^{\frac{KR^{\text{dist}}}{B}} \exp \left( \frac{1}{\overbar{\rho}^{\min}} \left( 2^{\frac{KR^{\text{dist}}}{B}} - 1 \right) \right) \\
	& < \frac{wcN}{\left(1 - \epsilon_l \right) \epsilon_l \lambda } \frac{1}{K^2} \ln \left( \frac{\lambda K + 1}{\left(1 - \epsilon_l \right) \epsilon_G \lambda} \right) - \frac{wcN}{\left(1 - \epsilon_l \right) \epsilon_l} \frac{1}{K}.
	\end{align}
	Otherwise stated, the optimality condition \eqref{eq:optimality_cond} cannot be satisfied for a value of $K$ that satisfies \eqref{eq:prop3_temp2}.
	Therefore, the optimal $K$, which minimizes the upper bound of the average completion time in the large data regime, should satisfy
	\begin{align}
	\nonumber &wN \frac{R^{\text{dist}} \ln 2 }{B} 2^{\frac{KR^{\text{dist}}}{B}} \exp \left( \frac{1}{\overbar{\rho}^{\min}} 2^{\frac{KR^{\text{dist}}}{B}} \right) \\
	&\geq \frac{wcN}{\left(1 - \epsilon_l \right) \epsilon_l \lambda } \frac{1}{K^2} \ln \left( \frac{\lambda K + 1}{\left(1 - \epsilon_l \right) \epsilon_G \lambda} \right) - \frac{wcN}{\left(1 - \epsilon_l \right) \epsilon_l} \frac{1}{K} \label{eq:prop3_temp1}
	\end{align}
	After some algebraic manipulations, \eqref{eq:prop3_temp1} can be simplified as \eqref{eq:nec_cond_ub}

\bibliographystyle{IEEEtran}
\bibliography{references}

\end{document}